\documentclass[reprint,amssymb,amsmath,aps,showpacs,10 pt,floatfix,prb,longbibliography]{revtex4-2}
\def\TS2{\mbox{$1T$-TaS$_2$}}
\def\cm{cm$^{-1}$}
\usepackage{graphicx}%
\usepackage{color}
\usepackage{epstopdf}
\usepackage{amssymb}
\usepackage{amsmath}
\usepackage{amsfonts}
\usepackage{xcolor}

\usepackage{color}
\usepackage[colorlinks,bookmarks=false,citecolor=darkblue,linkcolor=red,urlcolor=blue]{hyperref} 
\definecolor{darkred}{rgb}{0.7,0.0,0.0}

\definecolor{darkblue}{rgb}{0,0.02,0.45}

\definecolor{darkgreen}{rgb}{0.02,0.45,0.0}

\definecolor{violet}{rgb}{0.8,0.2,0.6}

\begin{document}

\preprint{Phys. Rev. Lett: 1T-TaS2}

\title{Unconventional anisotropic charge dynamics in bulk $1T$-TaS$_2$ \\ induced by interlayer dimerization}

\author{Achyut Tiwari}
\email{achyut.tiwari@pi1.uni-stuttgart.de}
\affiliation{1. Physikalisches Institut, Universit{\"a}t Stuttgart, Pfaffenwaldring 57, 70569 Stuttgart, Germany}
\author{Maxim Wenzel}
\affiliation{1. Physikalisches Institut, Universit{\"a}t Stuttgart, Pfaffenwaldring 57, 70569 Stuttgart, Germany}
\author{R. Mathew Roy}
\affiliation{1. Physikalisches Institut, Universit{\"a}t Stuttgart, Pfaffenwaldring 57, 70569 Stuttgart, Germany}
\author{Christian Prange}
\affiliation{1. Physikalisches Institut, Universit{\"a}t Stuttgart, Pfaffenwaldring 57, 70569 Stuttgart, Germany}
\author{Bruno Gompf}
\affiliation{1. Physikalisches Institut, Universit{\"a}t Stuttgart, Pfaffenwaldring 57, 70569 Stuttgart, Germany}
\author{Martin~Dressel}
%\email{dressel@pi1.physik.uni-stuttgart.de}
\affiliation{1. Physikalisches Institut, Universit{\"a}t Stuttgart, Pfaffenwaldring 57, 70569 Stuttgart, Germany}

\begin{abstract}

The commensurate charge-density-wave phase of the prototypical transition metal dichalcogenide $1T$-TaS$_2$ is investigated by temperature- and polarization-dependent infrared spectroscopy, revealing distinct charge dynamics parallel and perpendicular to the layers. Supported by density-functional-theory calculations, we show that the in-plane electronic structure in the low-temperature commensurate phase is reconstructed by the $\sqrt{13}\times\sqrt{13}$ distortion of the Ta layers. In contrast, the out-of-plane response is governed by a quasi-one-dimensional, Peierls-like dimerization of the two-dimensional Star-of-David layers. Our results identify this dimerization as the dominant mechanism of the metal-to-insulator transition in both directions, rather than correlation-driven Mott localization.

\end{abstract}

\maketitle

Layered quantum materials provide a natural platform for studying collective electronic phases, where the ground state is determined not only by the properties of individual atomic layers but also by interlayer coupling through van der Waals (vdW) and Coulomb interactions, as well as hybridization~\cite{novoselov2005two,geim2013van,Brotons2024}. Transition-metal dichalcogenides (TMDs) are a widely studied family in this class, hosting a variety of correlated and symmetry-broken phases and offering large tunability via temperature and pressure, which allows control of the interlayer degrees of freedom in natural heterostructures~\cite{Wilson1975,renjith4h,6Rpressure,Hu2026}. Additionally, twisted heterostructures enable access to regimes such as moiré-driven superconductivity and engineered heavy-fermion behavior~\cite{cao2018unconventional,vavno2021artificial}.

\TS2\ is a prototypical TMD characterized by a sequence of charge-density-wave (CDW) phases and pressure-induced superconductivity~\cite{Wilson1975,sipos2008mott}. Upon cooling, it undergoes transitions from an incommensurate to a nearly commensurate metallic CDW phase (NC-CDW) below $\sim 350$~K, and then into an insulating commensurate CDW (C-CDW) phase below $\sim 190$~K \cite{ThomsonPRB94}. The latter transition is first-order, exhibits a thermal hysteresis, and additionally proceeds through an intermediate triclinic state upon warming \cite{Manzke1989}. In the C-CDW phase, a $\sqrt{13}\times\sqrt{13}$ lattice distortion forms Star-of-David (SoD) clusters (see Fig.1), where 12 Ta-atoms constitute bonding pairs contributing to CDW gap, leaving half-filled central Ta-atom unpaired and localized \cite{FP&ET1979}. In th e absence of interlayer coupling, suppressed intralayer hopping renders the system a Mott insulator with localized $S = \frac{1}{2}$ spins on the triangular lattice of central Ta atoms, making \TS2\ a compelling candidate for realizing a quantum spin liquid ground state \cite{Kratochvilova2017,Klanjsek2017,Law2017}.

While the low-temperature insulating state in \TS2\ was originally attributed to purely two-dimensional electronic correlations and described as a Mott insulator \cite{tosatti1976, Kim1994}, emerging theoretical and experimental evidence has challenged this view. Recent calculations suggest that interlayer coupling plays a significant role in shaping the in-plane electronic structure, opening a gap even in the absence of strong on-site Coulomb interactions \cite{Ritschel2018, Lee2019, ritschel2015orbital}. Energy-dependent angle-resolved photoemission spectroscopy (ARPES) has further revealed finite dispersion along the out-of-plane direction \cite{Wang2020}. Meanwhile, scanning tunneling microscopy (STM) shows that the local spectrum is strongly dependent on termination and stacking \cite{Lee2023}, and advanced many-body theory suggests that correlation effects can coexist with a stacking-driven hybridization gap, particularly near surfaces or in locally perturbed regions \cite{Petocchi2022,Wang2024}.

Despite these advances, the bulk electronic response perpendicular to the planes remains poorly understood. ARPES and STM are inherently surface-sensitive and therefore do not directly probe the electrodynamics across multiple stacked layers. Recent transport measurements have revealed an unconventional resistivity anisotropy with higher conduction along the out-of-plane direction, in stark contrast to the typical behavior of quasi-two-dimensional materials \cite{Martino2020, Svetin2017}. Consistent with this, our recent ellipsometry study concluded that the transition is inherently three dimensional, revealing the pivotal role of interlayer coupling in the phase evolution \cite{Tiwari2025}. This highlights the importance of resolving the out-of-plane charge dynamics in order to uncover the true nature of the low-temperature ground state of \TS2\ and to advance the design of functional layered quantum materials. In particular, direct probes of the electronic response perpendicular to the layers are essential for disentangling the intertwined effects of stacking order and the reconstruction of the electronic structure due to the in-plane lattice distortion and consequently, to establish how interlayer interactions drive the phase transitions in these TMDs.

In this Letter, we probe the in-plane and out-of-plane electrodynamics of bulk \TS2\ using infrared spectroscopy over a broad energy range, across the Fermi level. We find an unusual anisotropic response in a vdW layered material, where the NC-CDW phase shows relatively metallic out-of-plane transport and reduced anisotropy.  The spectral-weight evolution is already strongly anisotropic in the NC-CDW regime, whereas below the C-CDW transition the gap energy shows distinct temperature dependences for the in-plane and out-of-plane directions. Combined with density-functional-theory (DFT) calculations, our findings suggest a Peierls-like interlayer dimerization responsible for the three-dimensional electronic reconstruction across the C-CDW phase, consistent with earlier evidence for the role of interlayer coupling in the phase evolution \cite{Tiwari2025}. Importantly, we demonstrate that this dimerization is the dominant mechanism behind the metal--insulator transition in bulk \TS2\, arguing against a purely Mott-localization-driven description.
%%%%%%%%%%%%%%%%%%%%%%%%%%%%%%%%%%%%%%%%%%%%%%%%%%%%%%%%%%%%%
\begin{figure}[!t]
	\centering
	\includegraphics[width=0.9\columnwidth]{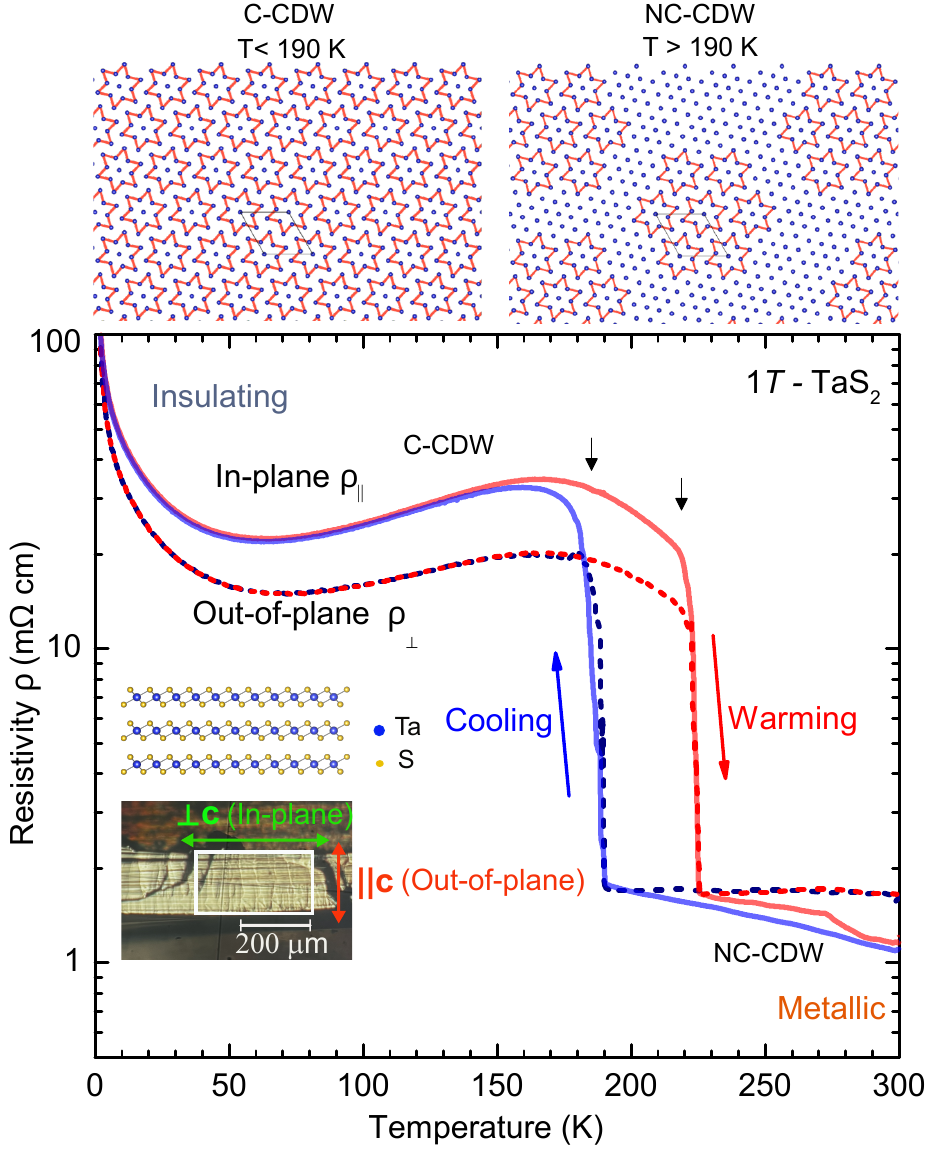}
	\caption{Temperature-dependent resistivity of a \TS2\ single crystal for in-plane (solid line) and out-of-plane (dashed line) upon cooling and heating. Black arrows mark the various CDW phase transition temperatures. Schematic illustrations (top) depict the sequence of nearly commensurate (NC-CDW) and commensurate (C-CDW) phases with decreasing temperature \cite{ThomsonPRB94}. The insets show the side view of crystal structure along with the ion-beam–polished cross-section of the \TS2\ single crystal used for the polarization-resolved optical measurements with green and orange labels indicating the polarization perpendicular and parallel to the $c$ axis, respectively. 
}	
	\label{fig:fig1}
\end{figure}
%%%%%%%%%%%%%%%%%%%%%%%%%%%%%%%%%%%%%%%%%%%%%%%%%%%%%%%%%%%%

 %%%%%%%%%%%%%%%%%%%%%%%%%%%%%%%%%%%%%%%%%%%%%%%%%%%%%%%%%%%%%
\begin{figure*}[!t]
	\centering
	\includegraphics[width=1.8\columnwidth]{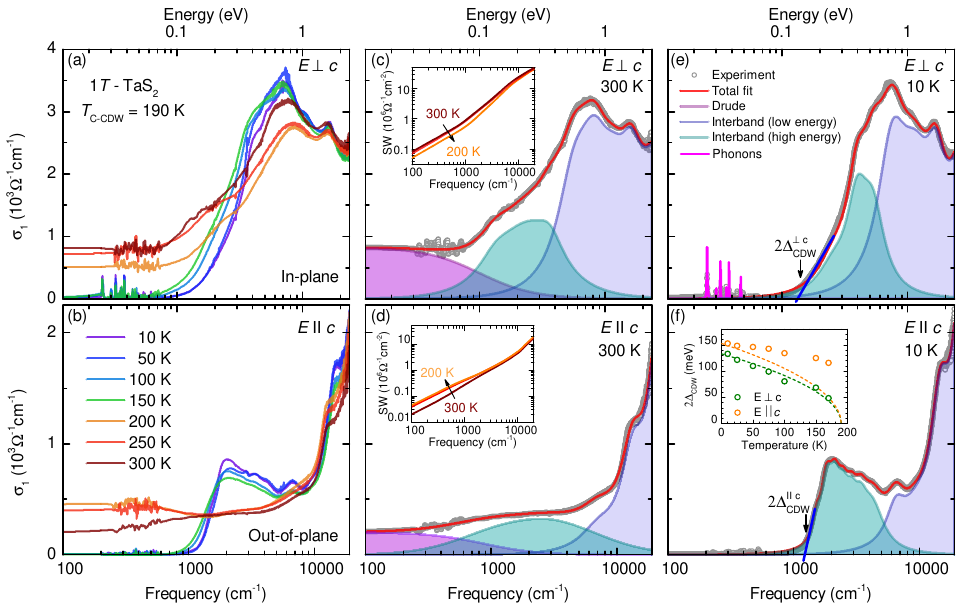}
	\caption{
(a, b) Real part of the optical conductivity $\sigma_1(\omega)$ of \TS2\ along the in-plane ($E \perp c$) and out-of-plane (\mbox{$E \parallel c$}) directions, calculated from the measured reflectivity via Kramers–Kronig analysis. (c–f) Decomposition of the optical conductivity into Drude, low energy interband transitions, high-energy interband transitions, and phonon modes for the in-plane (c, e), out-of-plane~(d, f) directions at $T = 300$ and 10~K, respectively. The optical spectra are modeled with the Drude-Lorentz approach as outlined in the Supplemental Material~\cite{SM}. In the C-CDW phase, the extrapolation of the steepest part of $\sigma_1(\omega)$ to the frequency axis is taken as the gap energy 2$\Delta_{\mathrm{CDW}}$ [see intersecting lines in panels (e) and (f)]. The insets of (c, d) show the temperature evolution of the frequency-dependent spectral weight (SW) in the NC-CDW phase. The inset of (f) shows the temperature evolution of the C-CDW gap obtained from this extrapolation of the absorption edge for in-plane and out-of-plane directions.}
	\label{fig:fig2}
\end{figure*}
%%%%%%%%%%%%%%%%%%%%%%%%%%%%%%%%%%%%%%%%%%%%%%%%%%%%%%%%%%%%

High-quality single crystals of \TS2\ were grown by chemical vapor transport and characterized using dc resistivity measurements. The polari\-zation-resolved reflectivity was measured on an ion-beam polished, optically flat crystal surfaces (see the inset of Fig.~\ref{fig:fig1}), over a broad frequency range of \mbox{200--19000~\cm} (25 meV--2.35~eV) at different temperatures down to 10~K. The optical conductivity was obtained via Kramers–Kronig analysis, with the full experimental details provided in the Supplemental Material~\cite{SM}.

 %%%%%%%%%%%%%%%%%%%%%%%%%%%%%%%%%%%%%%%%%%%%%%%%%%%%%%%%%%%%%
\begin{figure*}[!ht]
	\centering
	\includegraphics[width=1.7\columnwidth]{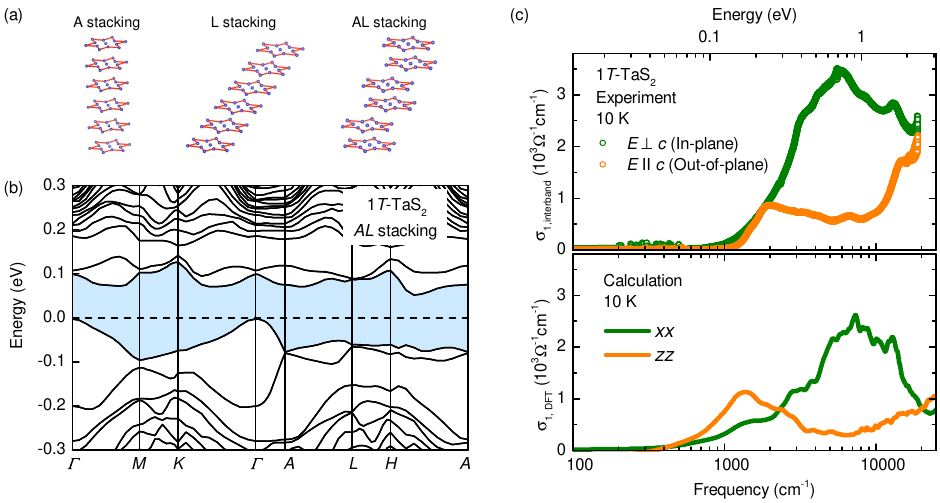}
	\caption{(a) Schematic illustration of stacking configurations A (aligned), L (ladder) and AL (alternating-ladder) of Star-of-David cluster along out-of-plane direction, only the Ta-atoms (blue) is shown for visualization purpose. (b) Calculated band structure for the low-temperature C-CDW phase for AL stacking with the shaded region highlights the gap at the Fermi level (c) Comparison of the measured real part of the optical conductivity $\sigma_1(\omega)$ (upper panel), with the optical response calculated from density-functional theory (lower panel) for the AL-stacked commensurate CDW structure. Note that a Gaussian broadening of 0.01~eV was applied to facilitate the comparison with the experiment.}
	\label{fig:fig3}
\end{figure*}
%%%%%%%%%%%%%%%%%%%%%%%%%%%%%%%%%%%%%%%%%%%%%%%%%%%%%%%%%%%%
 
Figure~\ref{fig:fig1} displays the temperature dependence of the in-plane ($\rho_{\parallel}$) and out-of-plane ($\rho_{\perp}$) resistivities of \TS2. Consistent with previous studies \cite{HAMBOURGER1980,TANI1981, MSarma1982}, $\rho_{\parallel}$ exhibits the nearly commensurate (NC) to commensurate (C) CDW transition as a first-order metal–insulator transition at $T_{\mathrm{C\text{-}CDW}}\approx 190$~K upon cooling, while upon heating it shows a broad hysteresis with an intermediate triclinic phase between 215~K and 275~K. Notably, $\rho_{\perp}$ exhibits a similar hysteresis and CDW transitions, with unusually low anisotropy for a layered TMD. This transport behavior points to an electronic reconstruction with a pronounced three-dimensional character, in agreement with earlier reports \cite{Svetin2017,Boix2021,Tiwari2025,Martino2020}. Intriguingly, in the \text{NC-CDW phase,} $\rho_{\parallel}$ increases upon cooling, whereas $\rho_{\perp}$ decreases, displaying a more conventional metallic behavior. This counter-intuitive trend indicates that nanoscale domains in the NC-CDW phase strongly suppress in-plane conduction \cite{Velebit2015}, whereas hopping due to interlayer coupling dominates the charge transport in the out-of-plane direction. 

The temperature-dependent real part of the optical conductivity, with the incident electric field polarized perpendicular and parallel to the $c$-axis is presented in Fig.~\ref{fig:fig2}(a) and (b), respectively. The in-plane ($\perp$c) optical response agrees well with earlier reports \cite{phonon2002, Lin2025}, corroborating the robustness of our measurements. In the NC-CDW phase, both in-plane and out-of-plane optical conductivites undergoes a weak Drude-like response below $\omega < 1000$~cm$^{-1}$. Interestingly, for the in-plane direction, the low-energy spectral weight, \mbox{$\mathrm{SW}=\int\sigma_{1}(\omega)\,\mathrm{d}\omega$,} decreases upon cooling [see inset of Fig.~2 (c)], resembling thermally activated behavior. In contrast, for the out-of-plane direction, the low-energy spectral weight increases [see inset of Fig.~2 (d)], reflecting the more conventional metallic behavior observed in the dc resistivity.

Upon cooling below $T_{\mathrm{C\text{-}CDW}}$, the optical conductivity shows a pronounced suppression  below $\sim$~1000~\cm, consistent with a gap opening and the onset of insulating behavior \cite{phonon2002}. In the in-plane direction, this is accompanied by an appearance of phonons at lower frequencies, as electronic screening is reduced. Apart from the expected sharpening of features upon cooling, the optical response exhibits a redistribution of spectral weight over a broad energy range, signaling a reconstruction of the electronic band structure across the transition, consistent with earlier ellipsometry results \cite{Tiwari2025}. In both directions, the suppressed low-energy spectral weight is transferred into the gap-region and higher-energy interband response. For the in-plane response, the total spectral weight remains nearly conserved within the measured range, whereas for the out-of-plane response the difference between the NC-CDW and C-CDW phases remains relatively modest and is comparable to the uncertainty associated with the high-energy extrapolation (see Sec.~S3 and Fig.~S4 of the Supplemental Material \cite{SM}). This analysis tracks the spectral-weight reshuffling associated with gap opening and band reconstruction, and does not suggest a correlation-driven transfer over much larger energy scales \cite{carbone2006plane}.

The gap energy, 2$\Delta_{\mathrm{CDW}}$, can be obtained by extra\-polating the steepest part of the absorption edge to zero, as illustrated by the solid lines in Fig.~\ref{fig:fig2}(e) and (f), yielding the values $\sim$~135 meV and $\sim$~153 meV at 10~K for the in-plane and out-of-plane direction, respectively. Alternatively, the zero crossing point of the difference optical conductivity leads to similar temperature dependence (see Sec.~S3 and Fig.~S5 of the supplemental material~\cite{SM}). The energy gap obtained for in-plane is consistent with the previous infrared, ARPES and STM studies, where the C-CDW gap was determined to fall in the range of \mbox{100--250~meV}  \cite{phonon2002, Wang2020, Wang2024,feng2023,Lin2025}. Moreover, the gap energy along the out-of-plane direction deduced from our optical measurements agrees well with the $\mathrm{k}_z$ bandwidth in the C-CDW phase observed by ARPES~\cite{Wang2020}.

The temperature evolution of the C-CDW gap energy is shown in the inset of Fig.~\ref{fig:fig2}(f). The in-plane gap exhibits a mean-field-like temperature dependence and can be described by \mbox{$2\Delta(T)\approx 2\Delta(0~\mathrm{K})\sqrt{1-\left(T/T_{\mathrm{CDW}}\right)}$}, where we assume $T_{\mathrm{CDW}}=190$~K and \mbox{$2\Delta^{\perp c}(0~\mathrm{K})=135$~meV.} Such mean-field-like gap evolution is a standard feature of density-waves and has been reported in optical studies of other CDW materials \cite{Grunerreview1988, MFPfuner2010,Perucchi2005}. In contrast, the out-of-plane gap evolves more abruptly and cannot be captured by the mean-field behavior \mbox{(2$\Delta^{\parallel{c}} (0~\mathrm{K}) = 153$ meV)}, indicating a distinct gap-formation linked to the out-of-plane electronic reconstruction. Similar deviations from mean-field gap evolution have been reported in SoD–based CDW systems, however, in those cases the anomaly appears for in-plane excitations \cite{EceCSV, He2024}. 
Notably, the ratio $2\Delta^{\perp c}(0)/k_{\mathrm{B}}T_{\mathrm{CDW}}\approx 8.2$  far exceed the weak-coupling BCS value $2\Delta(0)/k_{\mathrm{B}}T_c\simeq 3.52$, pointing to strong-coupling and fluctuation effects beyond a simple mean-field description.

Altogether, these findings demonstrate the intrinsically three-dimensional nature of the C-CDW ground state of \TS2, in line with our earlier ellipsometric results \cite{Tiwari2025}, and reveal a significant reconstruction of the low-energy electronic structure and the opening of energy gaps with distinct characteristics along the in-plane and out-of-plane directions. These results are consistent with recent theoretical and $k_z$-resolved ARPES studies, which render the low-temperature gap in bulk \TS2\ highly sensitive to the stacking of the SoD layers and the resulting interlayer hybridization \cite{Ritschel2018,Lee2019,Wang2020}. In particular, first-principle calculations propose that the metal-insulator transition may be driven by vertical ordering of the CDW layers, and a fully gapped Fermi surface can be realized for specific dimerized bilayer stackings even without including strong on-site interactions. Experimental confirmation, however, remains elusive due to thin samples, which often limits measurements to the $ab$-plane, and the inherently surface-sensitivity of many techniques, where information on the stacking sequence can only be inferred indirectly. In contrast, our optical measurements provide a bulk-sensitive and direct probe of both the in-plane and out-of-plane electronic structure. To interpret our findings within a theoretical framework, we perform DFT calculations of the electronic structure and optical conductivity for various stacking configurations [see Fig.~\ref{fig:fig3}(a) for a sketch of the aligned (A), ladder (L) and alternating-ladder (AL)] to distinguish the effects of the in-plane SoD distortion and the interlayer stacking.

 Consistent with previous studies, we find that the alternating-ladder (AL) stacking \cite{Wang2024} of the $\sqrt{13} \times \sqrt{13}$ CDW layers leads to a fully gapped electronic ground state, with a gap on the order of 120 meV as presented in Fig.~\ref{fig:fig3}(b). In contrast, other stacking configurations fail to reproduce the insulating ground state (see Figs.~S6 and S7 \cite{SM}). The remarkable agreement between the experimental and DFT-based interband optical conductivity for the AL stacking shown in Fig.~\ref{fig:fig3}(c) establishes that the gap-opening mechanism in bulk \TS2\ is a Peierls-like instability \cite{Peierls1955,Kennedy1987}, realized as interlayer dimerization of SoD layers that doubles the periodicity along the $c$-axis \cite{X-ray1984, Burri2025, Petocchi2022}. The out-of-plane optical conductivity is reproduced exclusively for AL stacking, whereas the in-plane interband response appears insensitive to the stacking order. In particular, using experimentally determined atomic parameters of the C-CDW state \cite{Petkov2022} leads to an excellent match between the experimental and DFT-based in-plane optical conductivities as presented in Fig.~S6(f) \cite{SM}. Such close agreement between the measured and calculated interband transitions argues against significant band-energy renormalization, which is commonly regarded as a hallmark of strong electronic correlations. These findings indicate that the in-plane $\sqrt{13} \times \sqrt{13}$ distortion drives the characteristic reconstruction of the in-plane electronic structure, only partially gapping the Fermi surface. This is consistent with related polytypes such as 4$H_b$-, \mbox{6$R$-TaS$_2$,} which undergo CDW transitions yet remain metallic \cite{renjith4h, PAL2023}. Subsequently, the interlayer dimerization, occurring alongside the in-plane SoD distortion, is the essential mechanism stabilizing the bulk insulating ground state.

The interplay between the in-plane SoD distortion and the out-of-plane dimerization also accounts for the observed anisotropic energy gap and its distinct temperature evolution. The out-of-plane gap is governed most directly by the stacking-dependent interlayer dimerization, whereas the in-plane gap reflects the combined effect of the intralayer CDW reconstruction and the interlayer hybridization that stabilizes the fully insulating bulk state. Thus, the in-plane gap evolves more smoothly, while the out-of-plane gap appears more abrupt across the first-order NC-CDW to C-CDW transition. The temperature dependence of the out-of-plane dc transport in the NC-CDW phase further supports that interlayer hopping is already relevant before the commensurate C-CDW transition. The resulting insulating state is therefore extremely sensitive to stacking, with interlayer hybridization providing the primary mechanism for bulk gap formation, without requiring a purely correlation-driven localization scenario. This is consistent with \citet{Petocchi2022}, who showed that hybridization and correlations can coexist but play distinct roles in the bulk. Within this framework, the termination dependence and the coexistence of different gap scales reported by STM can be understood as a consequence of local restacking and surface reconstruction induced by cleaving or interlayer sliding \cite{Lee2023,Wang2024,Origin2024}.

Furthermore, our findings do not exclude the presence of electron-electron interactions, however, they indicate that these are not the primary driver of the bulk insulating state. Moreover, comparison of the experimental Drude spectral weight with the DFT band value in the metallic NC-CDW phase provides a quantitative estimate of the correlation strength (see Sec.~S3 of the Supplemental Material \cite{SM}). The ratio between the experimental and band kinetic energies, $K_{\rm exp}/K_{\rm band}=\omega_{p,\rm exp}^2/\omega_{p,\rm band}^2$, is approximately $0.6$ for both polarizations, indicating only moderate correlations and placing bulk \TS2\ far from the Mott-insulating limit \cite{Qazilbash2009,Si2009,Ferber2010, Degiorgi2011}. The dominant gap-opening mechanism in the bulk electronic structure is instead a Peierls-like interlayer dimerization of SoD layers along the $c$-axis. These observations imply that the ground state in bulk \TS2\ is a three-dimensional band insulator, thereby challenging the Mott-insulating picture and, consequently, the realization of a quantum spin liquid phase \cite{Kratochvilova2017,Klanjsek2017}.

 Nevertheless, the pronounced sensitivity to stacking, supported by both theoretical and experimental findings, suggests a tangible route for manipulating materials: stacking faults, interlayer sliding, and optically or electrically induced restacking can selectively modulate interlayer hybridization, thereby tuning the electronic gap. This provides a foundation for engineering domain-wall metallicity, nonequilibrium phases, and flat-band regimes in specifically designed TaS$_2$ heterostructures  \cite{Hovden2016,Stahl2020,Liu2025,Bae2025}.

In conclusion, temperature- and polarization-resolved infrared spectroscopy, combined with DFT calculations, reveals the out-of-plane charge dynamics of \TS2\ across the first-order C-CDW transition. The anomalous dc anisotropy in the NC-CDW phase indicates enhanced interlayer hopping, deviating from the conventional quasi-two-dimensional behavior. In the C-CDW state, a clear optical gap develops, exhibiting a distinct temperature dependence along the in-plane and out-of-plane directions. This insulating ground state and the out-of-plane interband optical conductivity are reproduced by DFT only when interlayer dimerization is included. In contrast, the in-plane interband optical transitions appear less sensitive to stacking, suggesting that the primary reconstruction of the in-plane electronic structure is driven by the SoD distortion of the Ta layers. These results identify a quasi-one-dimensional, Peierls-like instability along the $c$-axis as the key ingredient stabilizing the bulk insulating state and establish interlayer stacking as an essential control parameter of the metal–insulator transition in \TS2, motivating targeted stacking engineering to tune correlated and nonequilibrium phases in layered CDW materials.

\textit{Acknowledgments}: We thank G. Untereiner and B. Fenk for technical support. We also thank T. Ritschel, W. Lu and L. Cao for providing atomic structures for the different stacking configurations. This work is supported by the Deutsche Forschungsgemeinschaft (DFG) under Grant No. DR228/63-1 and GO642/8-1.

\nocite{Tantar2010, Dressel2002, Quantum Espresso, Giannozzi2009, Giannozzi2017, wien2k, Blaha2020, Petkov2022, Draxl2006, Suh2000, osti_1192226, Ritschel2018}

\clearpage
\pagebreak

\onecolumngrid

\newpage
\vspace*{-2.5cm}
\hspace*{-2.5cm} {
  \centering
  \includegraphics[width=1.2\textwidth, page=1]{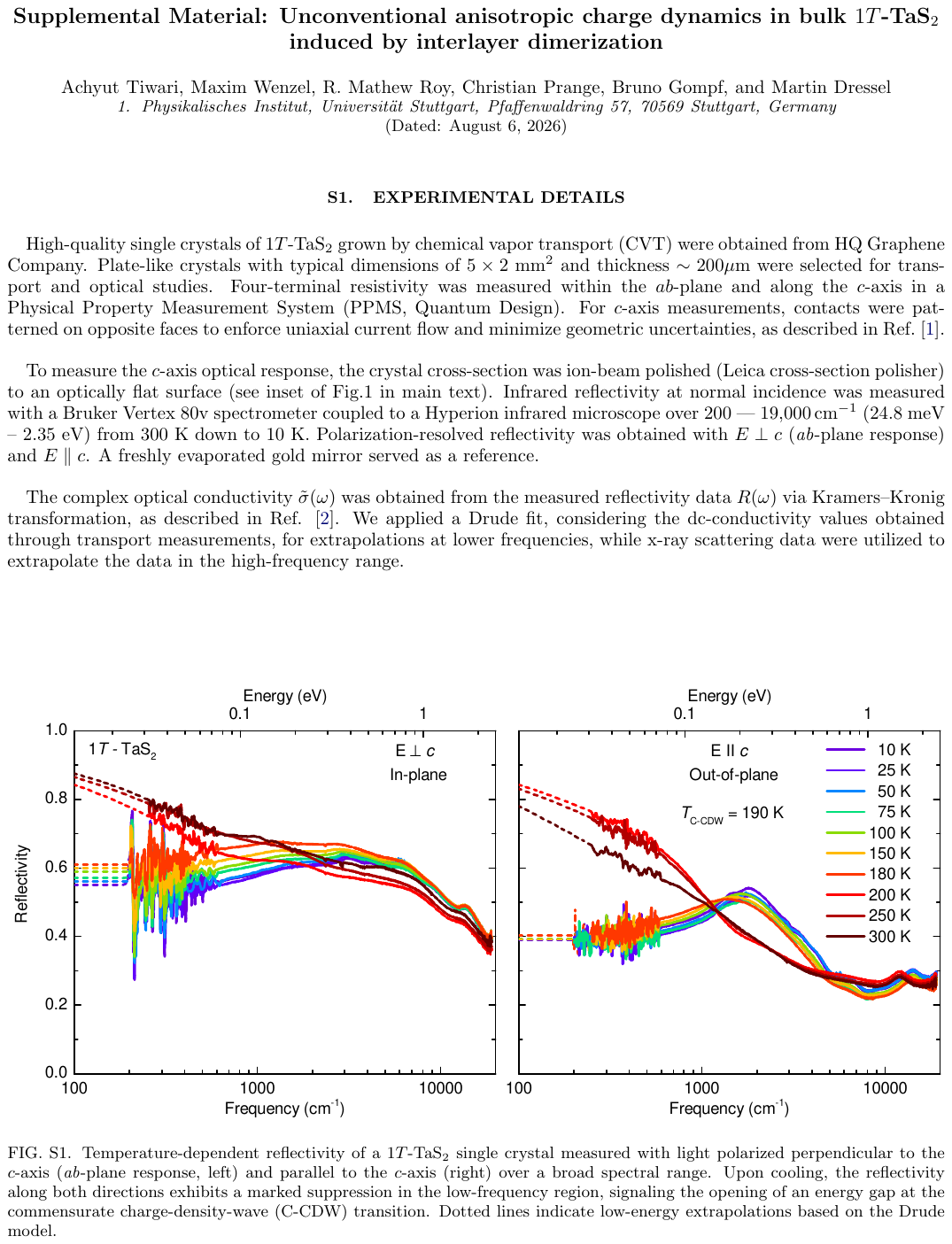} }
\vspace*{-2.5cm}
\hspace*{-2.5cm} {
  \centering
  \includegraphics[width=1.2\textwidth, page=2]{SM.pdf} }

\vspace*{-2.5cm}
\hspace*{-2.5cm} {
  \centering
  \includegraphics[width=1.2\textwidth, page=3]{SM.pdf} }

\vspace*{-2.5cm}
\hspace*{-2.5cm} {
  \centering
  \includegraphics[width=1.2\textwidth, page=4]{SM.pdf} }

\vspace*{-2.5cm}
\hspace*{-2.5cm} {
  \centering
  \includegraphics[width=1.2\textwidth, page=5]{SM.pdf} }

\vspace*{-2.5cm}
\hspace*{-2.5cm} {
  \centering
  \includegraphics[width=1.2\textwidth, page=6]{SM.pdf} }

\vspace*{0cm}
\hspace*{-2.5cm} {
  \centering
  \includegraphics[width=1.2\textwidth, page=7]{SM.pdf} }

\vspace*{0cm}
\hspace*{-2.5cm} {
  \centering
  \includegraphics[width=1.2\textwidth, page=8]{SM.pdf} }
\vspace*{-2.75cm}
\hspace*{-2.5cm}

\vspace*{0cm}
\hspace*{-2.5cm} {
  \centering
  \includegraphics[width=1.2\textwidth, page=9]{SM.pdf} }
\vspace*{-2.75cm}
\hspace*{-2.5cm}

\vspace*{0cm}
\hspace*{-2.5cm} {
  \centering
  \includegraphics[width=1.2\textwidth, page=10]{SM.pdf} }
\vspace*{-2.75cm}
\hspace*{-2.5cm}

\vspace*{0cm}
\hspace*{-2.5cm} {
  \centering
  \includegraphics[width=1.2\textwidth, page=11]{SM.pdf} }
\vspace*{-2.75cm}
\hspace*{-2.5cm}

\vspace*{0cm}
\hspace*{-2.5cm} {
  \centering
  \includegraphics[width=1.2\textwidth, page=12]{SM.pdf} }
\vspace*{-2.75cm}
\hspace*{-2.5cm}

\vspace*{0cm}
\hspace*{-2.5cm} {
  \centering
  \includegraphics[width=1.2\textwidth, page=13]{SM.pdf} }
\vspace*{-2.75cm}
\hspace*{-2.5cm}

\vspace*{0cm}
\hspace*{-2.5cm} {
  \centering
  \includegraphics[width=1.2\textwidth, page=14]{SM.pdf} }
\vspace*{-2.75cm}
\hspace*{-2.5cm}

\vspace*{0cm}
\hspace*{-2.5cm} {
  \centering
  \includegraphics[width=1.2\textwidth, page=15]{SM.pdf} }
\vspace*{-2.75cm}
\hspace*{-2.5cm}


\begin{thebibliography}{67}%
\makeatletter
\providecommand \@ifxundefined [1]{%
 \@ifx{#1\undefined}
}%
\providecommand \@ifnum [1]{%
 \ifnum #1\expandafter \@firstoftwo
 \else \expandafter \@secondoftwo
 \fi
}%
\providecommand \@ifx [1]{%
 \ifx #1\expandafter \@firstoftwo
 \else \expandafter \@secondoftwo
 \fi
}%
\providecommand \natexlab [1]{#1}%
\providecommand \enquote  [1]{``#1''}%
\providecommand \bibnamefont  [1]{#1}%
\providecommand \bibfnamefont [1]{#1}%
\providecommand \citenamefont [1]{#1}%
\providecommand \href@noop [0]{\@secondoftwo}%
\providecommand \href [0]{\begingroup \@sanitize@url \@href}%
\providecommand \@href[1]{\@@startlink{#1}\@@href}%
\providecommand \@@href[1]{\endgroup#1\@@endlink}%
\providecommand \@sanitize@url [0]{\catcode `\\12\catcode `\$12\catcode
  `\&12\catcode `\#12\catcode `\^12\catcode `\_12\catcode `\%12\relax}%
\providecommand \@@startlink[1]{}%
\providecommand \@@endlink[0]{}%
\providecommand \url  [0]{\begingroup\@sanitize@url \@url }%
\providecommand \@url [1]{\endgroup\@href {#1}{\urlprefix }}%
\providecommand \urlprefix  [0]{URL }%
\providecommand \Eprint [0]{\href }%
\providecommand \doibase [0]{https://doi.org/}%
\providecommand \selectlanguage [0]{\@gobble}%
\providecommand \bibinfo  [0]{\@secondoftwo}%
\providecommand \bibfield  [0]{\@secondoftwo}%
\providecommand \translation [1]{[#1]}%
\providecommand \BibitemOpen [0]{}%
\providecommand \bibitemStop [0]{}%
\providecommand \bibitemNoStop [0]{.\EOS\space}%
\providecommand \EOS [0]{\spacefactor3000\relax}%
\providecommand \BibitemShut  [1]{\csname bibitem#1\endcsname}%
\let\auto@bib@innerbib\@empty
%</preamble>
\bibitem [{\citenamefont {Novoselov}\ \emph {et~al.}(2005)\citenamefont
  {Novoselov}, \citenamefont {Jiang}, \citenamefont {Schedin}, \citenamefont
  {Booth}, \citenamefont {Khotkevich}, \citenamefont {Morozov},\ and\
  \citenamefont {Geim}}]{novoselov2005two}%
  \BibitemOpen
  \bibfield  {author} {\bibinfo {author} {\bibfnamefont {K.~S.}\ \bibnamefont
  {Novoselov}}, \bibinfo {author} {\bibfnamefont {D.}~\bibnamefont {Jiang}},
  \bibinfo {author} {\bibfnamefont {F.}~\bibnamefont {Schedin}}, \bibinfo
  {author} {\bibfnamefont {T.}~\bibnamefont {Booth}}, \bibinfo {author}
  {\bibfnamefont {V.}~\bibnamefont {Khotkevich}}, \bibinfo {author}
  {\bibfnamefont {S.}~\bibnamefont {Morozov}},\ and\ \bibinfo {author}
  {\bibfnamefont {A.~K.}\ \bibnamefont {Geim}},\ }\bibfield  {title} {\bibinfo
  {title} {Two-dimensional atomic crystals},\ }\href
  {https://doi.org/10.1073/pnas.0502848102} {\bibfield  {journal} {\bibinfo
  {journal} {Proc. Natl Acad. Sci.}\ }\textbf {\bibinfo {volume} {102}},\
  \bibinfo {pages} {10451} (\bibinfo {year} {2005})}\BibitemShut {NoStop}%
\bibitem [{\citenamefont {Geim}\ and\ \citenamefont
  {Grigorieva}(2013)}]{geim2013van}%
  \BibitemOpen
  \bibfield  {author} {\bibinfo {author} {\bibfnamefont {A.~K.}\ \bibnamefont
  {Geim}}\ and\ \bibinfo {author} {\bibfnamefont {I.~V.}\ \bibnamefont
  {Grigorieva}},\ }\bibfield  {title} {\bibinfo {title} {Van der waals
  heterostructures},\ }\href {https://doi.org/10.1038/nature12385} {\bibfield
  {journal} {\bibinfo  {journal} {Nature}\ }\textbf {\bibinfo {volume} {499}},\
  \bibinfo {pages} {419} (\bibinfo {year} {2013})}\BibitemShut {NoStop}%
\bibitem [{\citenamefont {Brotons-Gisbert}\ \emph {et~al.}(2024)\citenamefont
  {Brotons-Gisbert}, \citenamefont {Gerardot}, \citenamefont {Holleitner},\
  and\ \citenamefont {Wurstbauer}}]{Brotons2024}%
  \BibitemOpen
  \bibfield  {author} {\bibinfo {author} {\bibfnamefont {M.}~\bibnamefont
  {Brotons-Gisbert}}, \bibinfo {author} {\bibfnamefont {B.~D.}\ \bibnamefont
  {Gerardot}}, \bibinfo {author} {\bibfnamefont {A.~W.}\ \bibnamefont
  {Holleitner}},\ and\ \bibinfo {author} {\bibfnamefont {U.}~\bibnamefont
  {Wurstbauer}},\ }\bibfield  {title} {\bibinfo {title} {Interlayer and
  moir\'{e} excitons in atomically thin double layers: From individual quantum
  emitters to degenerate ensembles},\ }\href
  {https://doi.org/10.1557/s43577-024-00772-z} {\bibfield  {journal} {\bibinfo
  {journal} {MRS Bulletin}\ }\textbf {\bibinfo {volume} {49}},\ \bibinfo
  {pages} {914} (\bibinfo {year} {2024})}\BibitemShut {NoStop}%
\bibitem [{\citenamefont {Wilson}\ \emph {et~al.}(1975)\citenamefont {Wilson},
  \citenamefont {Salvo},\ and\ \citenamefont {Mahajan}}]{Wilson1975}%
  \BibitemOpen
  \bibfield  {author} {\bibinfo {author} {\bibfnamefont {J.}~\bibnamefont
  {Wilson}}, \bibinfo {author} {\bibfnamefont {F.~D.}\ \bibnamefont {Salvo}},\
  and\ \bibinfo {author} {\bibfnamefont {S.}~\bibnamefont {Mahajan}},\
  }\bibfield  {title} {\bibinfo {title} {Charge-density waves and superlattices
  in the metallic layered transition metal dichalcogenides},\ }\href
  {https://doi.org/10.1080/00018737500101391} {\bibfield  {journal} {\bibinfo
  {journal} {Adv. Phys.}\ }\textbf {\bibinfo {volume} {24}},\ \bibinfo {pages}
  {117} (\bibinfo {year} {1975})}\BibitemShut {NoStop}%
\bibitem [{\citenamefont {Mathew~Roy}\ \emph {et~al.}(2025)\citenamefont
  {Mathew~Roy}, \citenamefont {Feng}, \citenamefont {Wenzel}, \citenamefont
  {Hasse}, \citenamefont {Shekhar}, \citenamefont {Vergniory}, \citenamefont
  {Felser}, \citenamefont {Pronin},\ and\ \citenamefont {Dressel}}]{renjith4h}%
  \BibitemOpen
  \bibfield  {author} {\bibinfo {author} {\bibfnamefont {R.}~\bibnamefont
  {Mathew~Roy}}, \bibinfo {author} {\bibfnamefont {X.}~\bibnamefont {Feng}},
  \bibinfo {author} {\bibfnamefont {M.}~\bibnamefont {Wenzel}}, \bibinfo
  {author} {\bibfnamefont {V.}~\bibnamefont {Hasse}}, \bibinfo {author}
  {\bibfnamefont {C.}~\bibnamefont {Shekhar}}, \bibinfo {author} {\bibfnamefont
  {M.~G.}\ \bibnamefont {Vergniory}}, \bibinfo {author} {\bibfnamefont
  {C.}~\bibnamefont {Felser}}, \bibinfo {author} {\bibfnamefont {A.~V.}\
  \bibnamefont {Pronin}},\ and\ \bibinfo {author} {\bibfnamefont
  {M.}~\bibnamefont {Dressel}},\ }\bibfield  {title} {\bibinfo {title}
  {{Interlayer Charge Transfer Induced by Electronic Instabilities in the
  Natural van der Waals Heterostructure $4{H}_{b}$-${\text{TaS}}_{2}$}},\
  }\href {https://doi.org/10.1103/sgz5-qj71} {\bibfield  {journal} {\bibinfo
  {journal} {Phys. Rev. Lett.}\ }\textbf {\bibinfo {volume} {135}},\ \bibinfo
  {pages} {116503} (\bibinfo {year} {2025})}\BibitemShut {NoStop}%
\bibitem [{\citenamefont {Wang}\ \emph
  {et~al.}(2024{\natexlab{a}})\citenamefont {Wang}, \citenamefont {Han},
  \citenamefont {Sun}, \citenamefont {Wang}, \citenamefont {An}, \citenamefont
  {Chen}, \citenamefont {Zhang}, \citenamefont {Zhou}, \citenamefont {Zhou},\
  and\ \citenamefont {Yang}}]{6Rpressure}%
  \BibitemOpen
  \bibfield  {author} {\bibinfo {author} {\bibfnamefont {S.}~\bibnamefont
  {Wang}}, \bibinfo {author} {\bibfnamefont {Y.}~\bibnamefont {Han}}, \bibinfo
  {author} {\bibfnamefont {S.}~\bibnamefont {Sun}}, \bibinfo {author}
  {\bibfnamefont {S.}~\bibnamefont {Wang}}, \bibinfo {author} {\bibfnamefont
  {C.}~\bibnamefont {An}}, \bibinfo {author} {\bibfnamefont {C.}~\bibnamefont
  {Chen}}, \bibinfo {author} {\bibfnamefont {L.}~\bibnamefont {Zhang}},
  \bibinfo {author} {\bibfnamefont {Y.}~\bibnamefont {Zhou}}, \bibinfo {author}
  {\bibfnamefont {J.}~\bibnamefont {Zhou}},\ and\ \bibinfo {author}
  {\bibfnamefont {Z.}~\bibnamefont {Yang}},\ }\bibfield  {title} {\bibinfo
  {title} {{Pressure Induced Nonmonotonic Evolution of Superconductivity in
  $6R$-${\mathrm{TaS}}_{2}$ with a Natural Bulk Van der Waals
  Heterostructure}},\ }\href {https://doi.org/10.1103/PhysRevLett.133.056001}
  {\bibfield  {journal} {\bibinfo  {journal} {Phys. Rev. Lett.}\ }\textbf
  {\bibinfo {volume} {133}},\ \bibinfo {pages} {056001} (\bibinfo {year}
  {2024}{\natexlab{a}})}\BibitemShut {NoStop}%
\bibitem [{\citenamefont {Hu}\ \emph {et~al.}(2026)\citenamefont {Hu},
  \citenamefont {Li}, \citenamefont {Xu}, \citenamefont {Wu}, \citenamefont
  {Wu}, \citenamefont {Huang}, \citenamefont {Liu}, \citenamefont {Zhou},
  \citenamefont {Yuan}, \citenamefont {Wu}, \citenamefont {Dong}, \citenamefont
  {Hu}, \citenamefont {Weng},\ and\ \citenamefont {Wang}}]{Hu2026}%
  \BibitemOpen
  \bibfield  {author} {\bibinfo {author} {\bibfnamefont {T.}~\bibnamefont
  {Hu}}, \bibinfo {author} {\bibfnamefont {B.-X.}\ \bibnamefont {Li}}, \bibinfo
  {author} {\bibfnamefont {S.}~\bibnamefont {Xu}}, \bibinfo {author}
  {\bibfnamefont {S.}~\bibnamefont {Wu}}, \bibinfo {author} {\bibfnamefont
  {Q.}~\bibnamefont {Wu}}, \bibinfo {author} {\bibfnamefont {J.}~\bibnamefont
  {Huang}}, \bibinfo {author} {\bibfnamefont {Q.}~\bibnamefont {Liu}}, \bibinfo
  {author} {\bibfnamefont {X.}~\bibnamefont {Zhou}}, \bibinfo {author}
  {\bibfnamefont {J.}~\bibnamefont {Yuan}}, \bibinfo {author} {\bibfnamefont
  {D.}~\bibnamefont {Wu}}, \bibinfo {author} {\bibfnamefont {T.}~\bibnamefont
  {Dong}}, \bibinfo {author} {\bibfnamefont {J.}~\bibnamefont {Hu}}, \bibinfo
  {author} {\bibfnamefont {H.}~\bibnamefont {Weng}},\ and\ \bibinfo {author}
  {\bibfnamefont {N.}~\bibnamefont {Wang}},\ }\bibfield  {title} {\bibinfo
  {title} {{Charge density wave transition of pristine and organic-intercalated
  $1T$-${\mathrm{VSe}}_{2}$ studied by infrared spectroscopy}},\ }\bibfield
  {journal} {\bibinfo  {journal} {npj Quantum Mater.}\ }\href
  {https://doi.org/10.1038/s41535-026-00886-4} {10.1038/s41535-026-00886-4}
  (\bibinfo {year} {2026})\BibitemShut {NoStop}%
\bibitem [{\citenamefont {Cao}\ \emph {et~al.}(2018)\citenamefont {Cao},
  \citenamefont {Fatemi}, \citenamefont {Fang}, \citenamefont {Watanabe},
  \citenamefont {Taniguchi}, \citenamefont {Kaxiras},\ and\ \citenamefont
  {Jarillo-Herrero}}]{cao2018unconventional}%
  \BibitemOpen
  \bibfield  {author} {\bibinfo {author} {\bibfnamefont {Y.}~\bibnamefont
  {Cao}}, \bibinfo {author} {\bibfnamefont {V.}~\bibnamefont {Fatemi}},
  \bibinfo {author} {\bibfnamefont {S.}~\bibnamefont {Fang}}, \bibinfo {author}
  {\bibfnamefont {K.}~\bibnamefont {Watanabe}}, \bibinfo {author}
  {\bibfnamefont {T.}~\bibnamefont {Taniguchi}}, \bibinfo {author}
  {\bibfnamefont {E.}~\bibnamefont {Kaxiras}},\ and\ \bibinfo {author}
  {\bibfnamefont {P.}~\bibnamefont {Jarillo-Herrero}},\ }\bibfield  {title}
  {\bibinfo {title} {Unconventional superconductivity in magic-angle graphene
  superlattices},\ }\href {https://doi.org/10.1038/nature26160} {\bibfield
  {journal} {\bibinfo  {journal} {Nature}\ }\textbf {\bibinfo {volume} {556}},\
  \bibinfo {pages} {43} (\bibinfo {year} {2018})}\BibitemShut {NoStop}%
\bibitem [{\citenamefont {Va{\v{n}}o}\ \emph {et~al.}(2021)\citenamefont
  {Va{\v{n}}o}, \citenamefont {Amini}, \citenamefont {Ganguli}, \citenamefont
  {Chen}, \citenamefont {Lado}, \citenamefont {Kezilebieke},\ and\
  \citenamefont {Liljeroth}}]{vavno2021artificial}%
  \BibitemOpen
  \bibfield  {author} {\bibinfo {author} {\bibfnamefont {V.}~\bibnamefont
  {Va{\v{n}}o}}, \bibinfo {author} {\bibfnamefont {M.}~\bibnamefont {Amini}},
  \bibinfo {author} {\bibfnamefont {S.~C.}\ \bibnamefont {Ganguli}}, \bibinfo
  {author} {\bibfnamefont {G.}~\bibnamefont {Chen}}, \bibinfo {author}
  {\bibfnamefont {J.~L.}\ \bibnamefont {Lado}}, \bibinfo {author}
  {\bibfnamefont {S.}~\bibnamefont {Kezilebieke}},\ and\ \bibinfo {author}
  {\bibfnamefont {P.}~\bibnamefont {Liljeroth}},\ }\bibfield  {title} {\bibinfo
  {title} {Artificial heavy fermions in a van der waals heterostructure},\
  }\href {https://doi.org/10.1038/s41586-021-04021-0} {\bibfield  {journal}
  {\bibinfo  {journal} {Nature}\ }\textbf {\bibinfo {volume} {599}},\ \bibinfo
  {pages} {582} (\bibinfo {year} {2021})}\BibitemShut {NoStop}%
\bibitem [{\citenamefont {Sipos}\ \emph {et~al.}(2008)\citenamefont {Sipos},
  \citenamefont {Kusmartseva}, \citenamefont {Akrap}, \citenamefont {Berger},
  \citenamefont {Forr{\'o}},\ and\ \citenamefont
  {Tuti{\v{s}}}}]{sipos2008mott}%
  \BibitemOpen
  \bibfield  {author} {\bibinfo {author} {\bibfnamefont {B.}~\bibnamefont
  {Sipos}}, \bibinfo {author} {\bibfnamefont {A.~F.}\ \bibnamefont
  {Kusmartseva}}, \bibinfo {author} {\bibfnamefont {A.}~\bibnamefont {Akrap}},
  \bibinfo {author} {\bibfnamefont {H.}~\bibnamefont {Berger}}, \bibinfo
  {author} {\bibfnamefont {L.}~\bibnamefont {Forr{\'o}}},\ and\ \bibinfo
  {author} {\bibfnamefont {E.}~\bibnamefont {Tuti{\v{s}}}},\ }\bibfield
  {title} {\bibinfo {title} {{From Mott state to superconductivity in
  $1T$-${\mathrm{TaS}}_{2}$}},\ }\href {https://doi.org/10.1038/nmat2318}
  {\bibfield  {journal} {\bibinfo  {journal} {Nat. Mater.}\ }\textbf {\bibinfo
  {volume} {7}},\ \bibinfo {pages} {960} (\bibinfo {year} {2008})}\BibitemShut
  {NoStop}%
\bibitem [{\citenamefont {Thomson}\ \emph {et~al.}(1994)\citenamefont
  {Thomson}, \citenamefont {Burk}, \citenamefont {Zettl},\ and\ \citenamefont
  {Clarke}}]{ThomsonPRB94}%
  \BibitemOpen
  \bibfield  {author} {\bibinfo {author} {\bibfnamefont {R.~E.}\ \bibnamefont
  {Thomson}}, \bibinfo {author} {\bibfnamefont {B.}~\bibnamefont {Burk}},
  \bibinfo {author} {\bibfnamefont {A.}~\bibnamefont {Zettl}},\ and\ \bibinfo
  {author} {\bibfnamefont {J.}~\bibnamefont {Clarke}},\ }\bibfield  {title}
  {\bibinfo {title} {{Scanning tunneling microscopy of the charge-density-wave
  structure in $1T$-${\mathrm{TaS}}_{2}$}},\ }\href
  {https://doi.org/10.1103/PhysRevB.49.16899} {\bibfield  {journal} {\bibinfo
  {journal} {Phys. Rev. B}\ }\textbf {\bibinfo {volume} {49}},\ \bibinfo
  {pages} {16899} (\bibinfo {year} {1994})}\BibitemShut {NoStop}%
\bibitem [{\citenamefont {Manzke}\ \emph {et~al.}(1989)\citenamefont {Manzke},
  \citenamefont {Buslaps}, \citenamefont {Pfalzgraf}, \citenamefont
  {Skibowski},\ and\ \citenamefont {Anderson}}]{Manzke1989}%
  \BibitemOpen
  \bibfield  {author} {\bibinfo {author} {\bibfnamefont {R.}~\bibnamefont
  {Manzke}}, \bibinfo {author} {\bibfnamefont {T.}~\bibnamefont {Buslaps}},
  \bibinfo {author} {\bibfnamefont {B.}~\bibnamefont {Pfalzgraf}}, \bibinfo
  {author} {\bibfnamefont {M.}~\bibnamefont {Skibowski}},\ and\ \bibinfo
  {author} {\bibfnamefont {O.}~\bibnamefont {Anderson}},\ }\bibfield  {title}
  {\bibinfo {title} {{On the Phase Transitions in $1T$-${\mathrm{TaS}}_{2}$}},\
  }\href {https://doi.org/10.1209/0295-5075/8/2/015} {\bibfield  {journal}
  {\bibinfo  {journal} {Europhys. Lett.}\ }\textbf {\bibinfo {volume} {8}},\
  \bibinfo {pages} {195} (\bibinfo {year} {1989})}\BibitemShut {NoStop}%
\bibitem [{\citenamefont {Fazekas}\ and\ \citenamefont
  {Tosatti}(1979)}]{FP&ET1979}%
  \BibitemOpen
  \bibfield  {author} {\bibinfo {author} {\bibfnamefont {P.}~\bibnamefont
  {Fazekas}}\ and\ \bibinfo {author} {\bibfnamefont {E.}~\bibnamefont
  {Tosatti}},\ }\bibfield  {title} {\bibinfo {title} {{Electrical, structural
  and magnetic properties of pure and doped $1T$-${\mathrm{TaS}}_{2}$}},\
  }\href {https://doi.org/10.1080/13642817908245359} {\bibfield  {journal}
  {\bibinfo  {journal} {Philos. Mag. B}\ }\textbf {\bibinfo {volume} {39}},\
  \bibinfo {pages} {229} (\bibinfo {year} {1979})}\BibitemShut {NoStop}%
\bibitem [{\citenamefont {Kratochvilova}\ \emph {et~al.}(2017)\citenamefont
  {Kratochvilova}, \citenamefont {Hillier}, \citenamefont {Wildes},
  \citenamefont {Wang}, \citenamefont {Cheong},\ and\ \citenamefont
  {Park}}]{Kratochvilova2017}%
  \BibitemOpen
  \bibfield  {author} {\bibinfo {author} {\bibfnamefont {M.}~\bibnamefont
  {Kratochvilova}}, \bibinfo {author} {\bibfnamefont {A.~D.}\ \bibnamefont
  {Hillier}}, \bibinfo {author} {\bibfnamefont {A.~R.}\ \bibnamefont {Wildes}},
  \bibinfo {author} {\bibfnamefont {L.}~\bibnamefont {Wang}}, \bibinfo {author}
  {\bibfnamefont {S.-W.}\ \bibnamefont {Cheong}},\ and\ \bibinfo {author}
  {\bibfnamefont {J.-G.}\ \bibnamefont {Park}},\ }\bibfield  {title} {\bibinfo
  {title} {{The low-temperature highly correlated quantum phase in the
  charge-density-wave $1T$-${\mathrm{TaS}}_{2}$ compound}},\ }\href
  {https://doi.org/10.1038/s41535-017-0048-1} {\bibfield  {journal} {\bibinfo
  {journal} {npj Quantum Mater.}\ }\textbf {\bibinfo {volume} {2}},\ \bibinfo
  {pages} {42} (\bibinfo {year} {2017})}\BibitemShut {NoStop}%
\bibitem [{\citenamefont {Klanj\v{s}ek}\ \emph {et~al.}(2017)\citenamefont
  {Klanj\v{s}ek}, \citenamefont {Zorko}, \citenamefont {\v{Z}itko},
  \citenamefont {Mravlje}, \citenamefont {Jagli\v{c}i\'{c}}, \citenamefont
  {Biswas}, \citenamefont {Prelov\v{s}ek}, \citenamefont {Mihailovic},\ and\
  \citenamefont {Ar\v{c}on}}]{Klanjsek2017}%
  \BibitemOpen
  \bibfield  {author} {\bibinfo {author} {\bibfnamefont {M.}~\bibnamefont
  {Klanj\v{s}ek}}, \bibinfo {author} {\bibfnamefont {A.}~\bibnamefont {Zorko}},
  \bibinfo {author} {\bibfnamefont {R.}~\bibnamefont {\v{Z}itko}}, \bibinfo
  {author} {\bibfnamefont {J.}~\bibnamefont {Mravlje}}, \bibinfo {author}
  {\bibfnamefont {Z.}~\bibnamefont {Jagli\v{c}i\'{c}}}, \bibinfo {author}
  {\bibfnamefont {P.~K.}\ \bibnamefont {Biswas}}, \bibinfo {author}
  {\bibfnamefont {P.}~\bibnamefont {Prelov\v{s}ek}}, \bibinfo {author}
  {\bibfnamefont {D.}~\bibnamefont {Mihailovic}},\ and\ \bibinfo {author}
  {\bibfnamefont {D.}~\bibnamefont {Ar\v{c}on}},\ }\bibfield  {title} {\bibinfo
  {title} {A high-temperature quantum spin liquid with polaron spins},\ }\href
  {https://doi.org/10.1038/nphys4212} {\bibfield  {journal} {\bibinfo
  {journal} {Nat. Phys.}\ }\textbf {\bibinfo {volume} {13}},\ \bibinfo {pages}
  {1130} (\bibinfo {year} {2017})}\BibitemShut {NoStop}%
\bibitem [{\citenamefont {Law}\ and\ \citenamefont {Lee}(2017)}]{Law2017}%
  \BibitemOpen
  \bibfield  {author} {\bibinfo {author} {\bibfnamefont {K.~T.}\ \bibnamefont
  {Law}}\ and\ \bibinfo {author} {\bibfnamefont {P.~A.}\ \bibnamefont {Lee}},\
  }\bibfield  {title} {\bibinfo {title} {{$1T$-${\mathrm{TaS}}_{2}$ as a
  quantum spin liquid}},\ }\href {https://doi.org/10.1073/pnas.1706769114}
  {\bibfield  {journal} {\bibinfo  {journal} {Proc. Natl Acad. Sci. USA}\
  }\textbf {\bibinfo {volume} {114}},\ \bibinfo {pages} {6996} (\bibinfo {year}
  {2017})}\BibitemShut {NoStop}%
\bibitem [{\citenamefont {Tosatti}\ and\ \citenamefont
  {Fazekas}(1976)}]{tosatti1976}%
  \BibitemOpen
  \bibfield  {author} {\bibinfo {author} {\bibfnamefont {E.}~\bibnamefont
  {Tosatti}}\ and\ \bibinfo {author} {\bibfnamefont {P.}~\bibnamefont
  {Fazekas}},\ }\bibfield  {title} {\bibinfo {title} {{On the nature of the
  low-temperature phase of $1T$-${\mathrm{TaS}}_{2}$}},\ }\href
  {https://doi.org/10.1051/jphyscol:1976426} {\bibfield  {journal} {\bibinfo
  {journal} {{J. Phys. Colloq.}}\ }\textbf {\bibinfo {volume} {37}},\ \bibinfo
  {pages} {C4} (\bibinfo {year} {1976})}\BibitemShut {NoStop}%
\bibitem [{\citenamefont {Kim}\ \emph {et~al.}(1994)\citenamefont {Kim},
  \citenamefont {Yamaguchi}, \citenamefont {Hasegawa},\ and\ \citenamefont
  {Kitazawa}}]{Kim1994}%
  \BibitemOpen
  \bibfield  {author} {\bibinfo {author} {\bibfnamefont {J.-J.}\ \bibnamefont
  {Kim}}, \bibinfo {author} {\bibfnamefont {W.}~\bibnamefont {Yamaguchi}},
  \bibinfo {author} {\bibfnamefont {T.}~\bibnamefont {Hasegawa}},\ and\
  \bibinfo {author} {\bibfnamefont {K.}~\bibnamefont {Kitazawa}},\ }\bibfield
  {title} {\bibinfo {title} {{Observation of Mott Localization Gap Using Low
  Temperature Scanning Tunneling Spectroscopy in Commensurate
  $1T$-${\mathrm{TaS}}_{2}$}},\ }\href
  {https://doi.org/10.1103/PhysRevLett.73.2103} {\bibfield  {journal} {\bibinfo
   {journal} {Phys. Rev. Lett.}\ }\textbf {\bibinfo {volume} {73}},\ \bibinfo
  {pages} {2103} (\bibinfo {year} {1994})}\BibitemShut {NoStop}%
\bibitem [{\citenamefont {Ritschel}\ \emph {et~al.}(2018)\citenamefont
  {Ritschel}, \citenamefont {Berger},\ and\ \citenamefont
  {Geck}}]{Ritschel2018}%
  \BibitemOpen
  \bibfield  {author} {\bibinfo {author} {\bibfnamefont {T.}~\bibnamefont
  {Ritschel}}, \bibinfo {author} {\bibfnamefont {H.}~\bibnamefont {Berger}},\
  and\ \bibinfo {author} {\bibfnamefont {J.}~\bibnamefont {Geck}},\ }\bibfield
  {title} {\bibinfo {title} {{Stacking-driven gap formation in layered
  $1T$-${\mathrm{TaS}}_{2}$}},\ }\href
  {https://doi.org/10.1103/PhysRevB.98.195134} {\bibfield  {journal} {\bibinfo
  {journal} {Phys. Rev. B}\ }\textbf {\bibinfo {volume} {98}},\ \bibinfo
  {pages} {195134} (\bibinfo {year} {2018})}\BibitemShut {NoStop}%
\bibitem [{\citenamefont {Lee}\ \emph {et~al.}(2019)\citenamefont {Lee},
  \citenamefont {Goh},\ and\ \citenamefont {Cho}}]{Lee2019}%
  \BibitemOpen
  \bibfield  {author} {\bibinfo {author} {\bibfnamefont {S.-H.}\ \bibnamefont
  {Lee}}, \bibinfo {author} {\bibfnamefont {J.~S.}\ \bibnamefont {Goh}},\ and\
  \bibinfo {author} {\bibfnamefont {D.}~\bibnamefont {Cho}},\ }\bibfield
  {title} {\bibinfo {title} {{Origin of the Insulating Phase and First-Order
  Metal-Insulator Transition in $1T$-${\mathrm{TaS}}_{2}$}},\ }\href
  {https://doi.org/10.1103/PhysRevLett.122.106404} {\bibfield  {journal}
  {\bibinfo  {journal} {Phys. Rev. Lett.}\ }\textbf {\bibinfo {volume} {122}},\
  \bibinfo {pages} {106404} (\bibinfo {year} {2019})}\BibitemShut {NoStop}%
\bibitem [{\citenamefont {Ritschel}\ \emph {et~al.}(2015)\citenamefont
  {Ritschel}, \citenamefont {Trinckauf}, \citenamefont {Koepernik},
  \citenamefont {B{\"u}chner}, \citenamefont {Zimmermann}, \citenamefont
  {Berger}, \citenamefont {Joe}, \citenamefont {Abbamonte},\ and\ \citenamefont
  {Geck}}]{ritschel2015orbital}%
  \BibitemOpen
  \bibfield  {author} {\bibinfo {author} {\bibfnamefont {T.}~\bibnamefont
  {Ritschel}}, \bibinfo {author} {\bibfnamefont {J.}~\bibnamefont {Trinckauf}},
  \bibinfo {author} {\bibfnamefont {K.}~\bibnamefont {Koepernik}}, \bibinfo
  {author} {\bibfnamefont {B.}~\bibnamefont {B{\"u}chner}}, \bibinfo {author}
  {\bibfnamefont {M.~v.}\ \bibnamefont {Zimmermann}}, \bibinfo {author}
  {\bibfnamefont {H.}~\bibnamefont {Berger}}, \bibinfo {author} {\bibfnamefont
  {Y.}~\bibnamefont {Joe}}, \bibinfo {author} {\bibfnamefont {P.}~\bibnamefont
  {Abbamonte}},\ and\ \bibinfo {author} {\bibfnamefont {J.}~\bibnamefont
  {Geck}},\ }\bibfield  {title} {\bibinfo {title} {Orbital textures and charge
  density waves in transition metal dichalcogenides},\ }\href
  {https://doi.org/10.1038/nphys3267} {\bibfield  {journal} {\bibinfo
  {journal} {Nat. Phys.}\ }\textbf {\bibinfo {volume} {11}},\ \bibinfo {pages}
  {328} (\bibinfo {year} {2015})}\BibitemShut {NoStop}%
\bibitem [{\citenamefont {Wang}\ \emph {et~al.}(2020)\citenamefont {Wang},
  \citenamefont {Yao}, \citenamefont {Xin}, \citenamefont {Han}, \citenamefont
  {Wang}, \citenamefont {Chen}, \citenamefont {Cai}, \citenamefont {Li},\ and\
  \citenamefont {Zhang}}]{Wang2020}%
  \BibitemOpen
  \bibfield  {author} {\bibinfo {author} {\bibfnamefont {Y.~D.}\ \bibnamefont
  {Wang}}, \bibinfo {author} {\bibfnamefont {W.~L.}\ \bibnamefont {Yao}},
  \bibinfo {author} {\bibfnamefont {Z.~M.}\ \bibnamefont {Xin}}, \bibinfo
  {author} {\bibfnamefont {T.~T.}\ \bibnamefont {Han}}, \bibinfo {author}
  {\bibfnamefont {Z.~G.}\ \bibnamefont {Wang}}, \bibinfo {author}
  {\bibfnamefont {L.}~\bibnamefont {Chen}}, \bibinfo {author} {\bibfnamefont
  {C.}~\bibnamefont {Cai}}, \bibinfo {author} {\bibfnamefont {Y.}~\bibnamefont
  {Li}},\ and\ \bibinfo {author} {\bibfnamefont {Y.}~\bibnamefont {Zhang}},\
  }\bibfield  {title} {\bibinfo {title} {{Band insulator to Mott insulator
  transition in $1T$-${\mathrm{TaS}}_{2}$}},\ }\href
  {https://doi.org/10.1038/s41467-020-18040-4} {\bibfield  {journal} {\bibinfo
  {journal} {Nat. Commun.}\ }\textbf {\bibinfo {volume} {11}},\ \bibinfo
  {pages} {4215} (\bibinfo {year} {2020})}\BibitemShut {NoStop}%
\bibitem [{\citenamefont {Lee}\ and\ \citenamefont {Cho}(2023)}]{Lee2023}%
  \BibitemOpen
  \bibfield  {author} {\bibinfo {author} {\bibfnamefont {S.~H.}\ \bibnamefont
  {Lee}}\ and\ \bibinfo {author} {\bibfnamefont {D.}~\bibnamefont {Cho}},\
  }\bibfield  {title} {\bibinfo {title} {{Charge density wave surface
  reconstruction in a van der Waals layered material}},\ }\href
  {https://doi.org/10.1038/s41467-023-41500-6} {\bibfield  {journal} {\bibinfo
  {journal} {Nat. Commun.}\ }\textbf {\bibinfo {volume} {14}},\ \bibinfo
  {pages} {5735} (\bibinfo {year} {2023})}\BibitemShut {NoStop}%
\bibitem [{\citenamefont {Petocchi}\ \emph {et~al.}(2022)\citenamefont
  {Petocchi}, \citenamefont {Nicholson}, \citenamefont {Salzmann},
  \citenamefont {Pasquier}, \citenamefont {Yazyev}, \citenamefont {Monney},\
  and\ \citenamefont {Werner}}]{Petocchi2022}%
  \BibitemOpen
  \bibfield  {author} {\bibinfo {author} {\bibfnamefont {F.}~\bibnamefont
  {Petocchi}}, \bibinfo {author} {\bibfnamefont {C.~W.}\ \bibnamefont
  {Nicholson}}, \bibinfo {author} {\bibfnamefont {B.}~\bibnamefont {Salzmann}},
  \bibinfo {author} {\bibfnamefont {D.}~\bibnamefont {Pasquier}}, \bibinfo
  {author} {\bibfnamefont {O.~V.}\ \bibnamefont {Yazyev}}, \bibinfo {author}
  {\bibfnamefont {C.}~\bibnamefont {Monney}},\ and\ \bibinfo {author}
  {\bibfnamefont {P.}~\bibnamefont {Werner}},\ }\bibfield  {title} {\bibinfo
  {title} {{Mott versus Hybridization Gap in the Low-Temperature Phase of
  $1T$-${\mathrm{TaS}}_{2}$}},\ }\href
  {https://doi.org/10.1103/PhysRevLett.129.016402} {\bibfield  {journal}
  {\bibinfo  {journal} {Phys. Rev. Lett.}\ }\textbf {\bibinfo {volume} {129}},\
  \bibinfo {pages} {016402} (\bibinfo {year} {2022})}\BibitemShut {NoStop}%
\bibitem [{\citenamefont {Wang}\ \emph
  {et~al.}(2024{\natexlab{b}})\citenamefont {Wang}, \citenamefont {Li},
  \citenamefont {Luo}, \citenamefont {Gao}, \citenamefont {Han}, \citenamefont
  {Jiang}, \citenamefont {Tang}, \citenamefont {Ju}, \citenamefont {Li},
  \citenamefont {Lv}, \citenamefont {Cui}, \citenamefont {Yang}, \citenamefont
  {Sun}, \citenamefont {Zhu}, \citenamefont {Gao}, \citenamefont {Lu},
  \citenamefont {Sun}, \citenamefont {Xu}, \citenamefont {Xiong},\ and\
  \citenamefont {Cao}}]{Wang2024}%
  \BibitemOpen
  \bibfield  {author} {\bibinfo {author} {\bibfnamefont {Y.}~\bibnamefont
  {Wang}}, \bibinfo {author} {\bibfnamefont {Z.}~\bibnamefont {Li}}, \bibinfo
  {author} {\bibfnamefont {X.}~\bibnamefont {Luo}}, \bibinfo {author}
  {\bibfnamefont {J.}~\bibnamefont {Gao}}, \bibinfo {author} {\bibfnamefont
  {Y.}~\bibnamefont {Han}}, \bibinfo {author} {\bibfnamefont {J.}~\bibnamefont
  {Jiang}}, \bibinfo {author} {\bibfnamefont {J.}~\bibnamefont {Tang}},
  \bibinfo {author} {\bibfnamefont {H.}~\bibnamefont {Ju}}, \bibinfo {author}
  {\bibfnamefont {T.}~\bibnamefont {Li}}, \bibinfo {author} {\bibfnamefont
  {R.}~\bibnamefont {Lv}}, \bibinfo {author} {\bibfnamefont {S.}~\bibnamefont
  {Cui}}, \bibinfo {author} {\bibfnamefont {Y.}~\bibnamefont {Yang}}, \bibinfo
  {author} {\bibfnamefont {Y.}~\bibnamefont {Sun}}, \bibinfo {author}
  {\bibfnamefont {J.}~\bibnamefont {Zhu}}, \bibinfo {author} {\bibfnamefont
  {X.}~\bibnamefont {Gao}}, \bibinfo {author} {\bibfnamefont {W.}~\bibnamefont
  {Lu}}, \bibinfo {author} {\bibfnamefont {Z.}~\bibnamefont {Sun}}, \bibinfo
  {author} {\bibfnamefont {H.}~\bibnamefont {Xu}}, \bibinfo {author}
  {\bibfnamefont {Y.}~\bibnamefont {Xiong}},\ and\ \bibinfo {author}
  {\bibfnamefont {L.}~\bibnamefont {Cao}},\ }\bibfield  {title} {\bibinfo
  {title} {{Dualistic insulator states in $1T$-${\mathrm{TaS}}_{2}$
  crystals}},\ }\href {https://doi.org/10.1038/s41467-024-47728-0} {\bibfield
  {journal} {\bibinfo  {journal} {Nat. Commun.}\ }\textbf {\bibinfo {volume}
  {15}},\ \bibinfo {pages} {3425} (\bibinfo {year}
  {2024}{\natexlab{b}})}\BibitemShut {NoStop}%
\bibitem [{\citenamefont {Martino}\ \emph {et~al.}(2020)\citenamefont
  {Martino}, \citenamefont {Pisoni}, \citenamefont {Ćirić}, \citenamefont
  {Arakcheeva}, \citenamefont {Berger}, \citenamefont {Akrap}, \citenamefont
  {Putzke}, \citenamefont {Moll}, \citenamefont {Batistić}, \citenamefont
  {Tutiš}, \citenamefont {Forró},\ and\ \citenamefont
  {Semeniuk}}]{Martino2020}%
  \BibitemOpen
  \bibfield  {author} {\bibinfo {author} {\bibfnamefont {E.}~\bibnamefont
  {Martino}}, \bibinfo {author} {\bibfnamefont {A.}~\bibnamefont {Pisoni}},
  \bibinfo {author} {\bibfnamefont {L.}~\bibnamefont {Ćirić}}, \bibinfo
  {author} {\bibfnamefont {A.}~\bibnamefont {Arakcheeva}}, \bibinfo {author}
  {\bibfnamefont {H.}~\bibnamefont {Berger}}, \bibinfo {author} {\bibfnamefont
  {A.}~\bibnamefont {Akrap}}, \bibinfo {author} {\bibfnamefont
  {C.}~\bibnamefont {Putzke}}, \bibinfo {author} {\bibfnamefont {P.~J.}\
  \bibnamefont {Moll}}, \bibinfo {author} {\bibfnamefont {I.}~\bibnamefont
  {Batistić}}, \bibinfo {author} {\bibfnamefont {E.}~\bibnamefont {Tutiš}},
  \bibinfo {author} {\bibfnamefont {L.}~\bibnamefont {Forró}},\ and\ \bibinfo
  {author} {\bibfnamefont {K.}~\bibnamefont {Semeniuk}},\ }\bibfield  {title}
  {\bibinfo {title} {{Preferential out-of-plane conduction and
  quasi-one-dimensional electronic states in layered
  $1T$-${\mathrm{TaS}}_{2}$}},\ }\href
  {https://doi.org/10.1038/s41699-020-0145-z} {\bibfield  {journal} {\bibinfo
  {journal} {npj 2D Mater. Appl.}\ }\textbf {\bibinfo {volume} {4}},\ \bibinfo
  {pages} {7} (\bibinfo {year} {2020})}\BibitemShut {NoStop}%
\bibitem [{\citenamefont {Svetin}\ \emph {et~al.}(2017)\citenamefont {Svetin},
  \citenamefont {Vaskivskyi}, \citenamefont {Brazovskii},\ and\ \citenamefont
  {Mihailovic}}]{Svetin2017}%
  \BibitemOpen
  \bibfield  {author} {\bibinfo {author} {\bibfnamefont {D.}~\bibnamefont
  {Svetin}}, \bibinfo {author} {\bibfnamefont {I.}~\bibnamefont {Vaskivskyi}},
  \bibinfo {author} {\bibfnamefont {S.}~\bibnamefont {Brazovskii}},\ and\
  \bibinfo {author} {\bibfnamefont {D.}~\bibnamefont {Mihailovic}},\ }\bibfield
   {title} {\bibinfo {title} {{Three-dimensional resistivity and switching
  between correlated electronic states in $1T$-${\mathrm{TaS}}_{2}$}},\ }\href
  {https://doi.org/10.1038/srep46048} {\bibfield  {journal} {\bibinfo
  {journal} {Sci. Rep.}\ }\textbf {\bibinfo {volume} {7}},\ \bibinfo {pages}
  {46048} (\bibinfo {year} {2017})}\BibitemShut {NoStop}%
\bibitem [{\citenamefont {Tiwari}\ \emph {et~al.}(2026)\citenamefont {Tiwari},
  \citenamefont {Gompf},\ and\ \citenamefont {Dressel}}]{Tiwari2025}%
  \BibitemOpen
  \bibfield  {author} {\bibinfo {author} {\bibfnamefont {A.}~\bibnamefont
  {Tiwari}}, \bibinfo {author} {\bibfnamefont {B.}~\bibnamefont {Gompf}},\ and\
  \bibinfo {author} {\bibfnamefont {M.}~\bibnamefont {Dressel}},\ }\bibfield
  {title} {\bibinfo {title} {{Interlayer coupling driven phase evolution in
  hyperbolic $1T$-${\mathrm{TaS}}_{2}$ revealed by spectroscopic
  ellipsometry}},\ }\href {https://doi.org/10.1063/5.0315654} {\bibfield
  {journal} {\bibinfo  {journal} {Appl. Phys. Lett.}\ }\textbf {\bibinfo
  {volume} {128}},\ \bibinfo {pages} {063104} (\bibinfo {year}
  {2026})}\BibitemShut {NoStop}%
\bibitem [{SM()}]{SM}%
  \BibitemOpen
  \href@noop {} {\bibinfo {title} {{See Supplemental Material at [URL] for
  experimental details, data analysis, details on DFT calculations, and
  additional computational results, which includes Refs. \cite{Tantar2010,
  Dressel2002, Giannozzi2009, Giannozzi2017, wien2k, Blaha2020, Petkov2022,
  Draxl2006, Suh2000, osti_1192226, Ritschel2018, Wang2024}}}}\BibitemShut
  {NoStop}%
\bibitem [{\citenamefont {Hambourger}\ and\ \citenamefont {{Di
  Salvo}}(1980)}]{HAMBOURGER1980}%
  \BibitemOpen
  \bibfield  {author} {\bibinfo {author} {\bibfnamefont {P.}~\bibnamefont
  {Hambourger}}\ and\ \bibinfo {author} {\bibfnamefont {F.}~\bibnamefont {{Di
  Salvo}}},\ }\bibfield  {title} {\bibinfo {title} {{Transport properties of
  $1T$-${\mathrm{TaS}}_{2-x}{\mathrm{Se}}_{x}$}},\ }\href
  {https://doi.org/https://doi.org/10.1016/0038-1098(80)90169-6} {\bibfield
  {journal} {\bibinfo  {journal} {Solid State Commun.}\ }\textbf {\bibinfo
  {volume} {35}},\ \bibinfo {pages} {405} (\bibinfo {year} {1980})}\BibitemShut
  {NoStop}%
\bibitem [{\citenamefont {Tani}\ \emph {et~al.}(1981)\citenamefont {Tani},
  \citenamefont {Okajima}, \citenamefont {Itoh},\ and\ \citenamefont
  {Tanaka}}]{TANI1981}%
  \BibitemOpen
  \bibfield  {author} {\bibinfo {author} {\bibfnamefont {T.}~\bibnamefont
  {Tani}}, \bibinfo {author} {\bibfnamefont {K.}~\bibnamefont {Okajima}},
  \bibinfo {author} {\bibfnamefont {T.}~\bibnamefont {Itoh}},\ and\ \bibinfo
  {author} {\bibfnamefont {S.}~\bibnamefont {Tanaka}},\ }\bibfield  {title}
  {\bibinfo {title} {{Electronic transport properties in
  $1T$-${\mathrm{TaS}}_{2}$}},\ }\href
  {https://doi.org/https://doi.org/10.1016/0378-4363(81)90230-8} {\bibfield
  {journal} {\bibinfo  {journal} {Physica B+C}\ }\textbf {\bibinfo {volume}
  {105}},\ \bibinfo {pages} {127} (\bibinfo {year} {1981})}\BibitemShut
  {NoStop}%
\bibitem [{\citenamefont {Sarma}\ \emph {et~al.}(1982)\citenamefont {Sarma},
  \citenamefont {Beal}, \citenamefont {Nulsen},\ and\ \citenamefont
  {Friend}}]{MSarma1982}%
  \BibitemOpen
  \bibfield  {author} {\bibinfo {author} {\bibfnamefont {M.}~\bibnamefont
  {Sarma}}, \bibinfo {author} {\bibfnamefont {A.~R.}\ \bibnamefont {Beal}},
  \bibinfo {author} {\bibfnamefont {S.}~\bibnamefont {Nulsen}},\ and\ \bibinfo
  {author} {\bibfnamefont {R.~H.}\ \bibnamefont {Friend}},\ }\bibfield  {title}
  {\bibinfo {title} {{Transport and optical properties of the hydrazine
  intercalation complexes of $1T$-${\mathrm{TaS}}_{2}$}},\ }\href
  {https://doi.org/10.1088/0022-3719/15/3/014} {\bibfield  {journal} {\bibinfo
  {journal} {J. Phys. C: Solid State Phys.}\ }\textbf {\bibinfo {volume}
  {15}},\ \bibinfo {pages} {477} (\bibinfo {year} {1982})}\BibitemShut
  {NoStop}%
\bibitem [{\citenamefont {Boix-Constant}\ \emph {et~al.}(2021)\citenamefont
  {Boix-Constant}, \citenamefont {Ma{\~n}as-Valero}, \citenamefont
  {C{\'o}rdoba}, \citenamefont {Baldov{\'i}}, \citenamefont {Rubio},\ and\
  \citenamefont {Coronado}}]{Boix2021}%
  \BibitemOpen
  \bibfield  {author} {\bibinfo {author} {\bibfnamefont {C.}~\bibnamefont
  {Boix-Constant}}, \bibinfo {author} {\bibfnamefont {S.}~\bibnamefont
  {Ma{\~n}as-Valero}}, \bibinfo {author} {\bibfnamefont {R.}~\bibnamefont
  {C{\'o}rdoba}}, \bibinfo {author} {\bibfnamefont {J.~J.}\ \bibnamefont
  {Baldov{\'i}}}, \bibinfo {author} {\bibfnamefont {{\'A}.}~\bibnamefont
  {Rubio}},\ and\ \bibinfo {author} {\bibfnamefont {E.}~\bibnamefont
  {Coronado}},\ }\bibfield  {title} {\bibinfo {title} {{Out-of-plane transport
  of $1T$-${\mathrm{TaS}}_{2}$/graphene-based van der Waals
  heterostructures}},\ }\href {https://doi.org/10.1021/acsnano.1c03012}
  {\bibfield  {journal} {\bibinfo  {journal} {ACS Nano}\ }\textbf {\bibinfo
  {volume} {15}},\ \bibinfo {pages} {11898} (\bibinfo {year}
  {2021})}\BibitemShut {NoStop}%
\bibitem [{\citenamefont {Velebit}(2015)}]{Velebit2015}%
  \BibitemOpen
  \bibfield  {author} {\bibinfo {author} {\bibfnamefont {K.}~\bibnamefont
  {Velebit}},\ }\emph {\bibinfo {title} {Effects of superstructuring on optical
  and transport properties of selected layered materials}},\ \href@noop {}
  {Ph.D. thesis},\ \bibinfo  {school} {University of Zagreb} (\bibinfo {year}
  {2015})\BibitemShut {NoStop}%
\bibitem [{\citenamefont {Gasparov}\ \emph {et~al.}(2002)\citenamefont
  {Gasparov}, \citenamefont {Brown}, \citenamefont {Wint}, \citenamefont
  {Tanner}, \citenamefont {Berger}, \citenamefont {Margaritondo}, \citenamefont
  {Ga\'al},\ and\ \citenamefont {Forr\'o}}]{phonon2002}%
  \BibitemOpen
  \bibfield  {author} {\bibinfo {author} {\bibfnamefont {L.~V.}\ \bibnamefont
  {Gasparov}}, \bibinfo {author} {\bibfnamefont {K.~G.}\ \bibnamefont {Brown}},
  \bibinfo {author} {\bibfnamefont {A.~C.}\ \bibnamefont {Wint}}, \bibinfo
  {author} {\bibfnamefont {D.~B.}\ \bibnamefont {Tanner}}, \bibinfo {author}
  {\bibfnamefont {H.}~\bibnamefont {Berger}}, \bibinfo {author} {\bibfnamefont
  {G.}~\bibnamefont {Margaritondo}}, \bibinfo {author} {\bibfnamefont
  {R.}~\bibnamefont {Ga\'al}},\ and\ \bibinfo {author} {\bibfnamefont
  {L.}~\bibnamefont {Forr\'o}},\ }\bibfield  {title} {\bibinfo {title} {{Phonon
  anomaly at the charge ordering transition in $1T$-${\mathrm{TaS}}_{2}$}},\
  }\href {https://doi.org/10.1103/PhysRevB.66.094301} {\bibfield  {journal}
  {\bibinfo  {journal} {Phys. Rev. B}\ }\textbf {\bibinfo {volume} {66}},\
  \bibinfo {pages} {094301} (\bibinfo {year} {2002})}\BibitemShut {NoStop}%
\bibitem [{\citenamefont {Lin}\ \emph {et~al.}(2025)\citenamefont {Lin},
  \citenamefont {Li}, \citenamefont {Cao}, \citenamefont {Gao}, \citenamefont
  {Luo}, \citenamefont {Sun}, \citenamefont {Lu}, \citenamefont {Wang},
  \citenamefont {Guo},\ and\ \citenamefont {Zhu}}]{Lin2025}%
  \BibitemOpen
  \bibfield  {author} {\bibinfo {author} {\bibfnamefont {Z.}~\bibnamefont
  {Lin}}, \bibinfo {author} {\bibfnamefont {J.}~\bibnamefont {Li}}, \bibinfo
  {author} {\bibfnamefont {X.}~\bibnamefont {Cao}}, \bibinfo {author}
  {\bibfnamefont {J.}~\bibnamefont {Gao}}, \bibinfo {author} {\bibfnamefont
  {X.}~\bibnamefont {Luo}}, \bibinfo {author} {\bibfnamefont {Y.}~\bibnamefont
  {Sun}}, \bibinfo {author} {\bibfnamefont {Y.}~\bibnamefont {Lu}}, \bibinfo
  {author} {\bibfnamefont {N.}~\bibnamefont {Wang}}, \bibinfo {author}
  {\bibfnamefont {J.}~\bibnamefont {Guo}},\ and\ \bibinfo {author}
  {\bibfnamefont {X.}~\bibnamefont {Zhu}},\ }\bibfield  {title} {\bibinfo
  {title} {{Interlayer hopping between a surface Mott insulator and a bulk band
  insulator in layered $1T$-${\mathrm{TaS}}_{2}$}},\ }\href
  {https://doi.org/10.1103/PhysRevB.111.075126} {\bibfield  {journal} {\bibinfo
   {journal} {Phys. Rev. B}\ }\textbf {\bibinfo {volume} {111}},\ \bibinfo
  {pages} {075126} (\bibinfo {year} {2025})}\BibitemShut {NoStop}%
\bibitem [{\citenamefont {Carbone}\ \emph {et~al.}(2006)\citenamefont
  {Carbone}, \citenamefont {Kuzmenko}, \citenamefont {Molegraaf}, \citenamefont
  {van Heumen}, \citenamefont {Giannini},\ and\ \citenamefont {van~der
  Marel}}]{carbone2006plane}%
  \BibitemOpen
  \bibfield  {author} {\bibinfo {author} {\bibfnamefont {F.}~\bibnamefont
  {Carbone}}, \bibinfo {author} {\bibfnamefont {A.~B.}\ \bibnamefont
  {Kuzmenko}}, \bibinfo {author} {\bibfnamefont {H.~J.~A.}\ \bibnamefont
  {Molegraaf}}, \bibinfo {author} {\bibfnamefont {E.}~\bibnamefont {van
  Heumen}}, \bibinfo {author} {\bibfnamefont {E.}~\bibnamefont {Giannini}},\
  and\ \bibinfo {author} {\bibfnamefont {D.}~\bibnamefont {van~der Marel}},\
  }\bibfield  {title} {\bibinfo {title} {{In-plane optical spectral weight
  transfer in optimally doped
  ${\mathrm{Bi}}_{2}{\mathrm{Sr}}_{2}{\mathrm{Ca}}_{2}{\mathrm{Cu}}_{3}{\mathrm{O}}_{10}$}},\
  }\href {https://doi.org/10.1103/PhysRevB.74.024502} {\bibfield  {journal}
  {\bibinfo  {journal} {Phys. Rev. B}\ }\textbf {\bibinfo {volume} {74}},\
  \bibinfo {pages} {024502} (\bibinfo {year} {2006})}\BibitemShut {NoStop}%
\bibitem [{\citenamefont {Feng}\ \emph {et~al.}(2023)\citenamefont {Feng},
  \citenamefont {Farrar}, \citenamefont {Sayers}, \citenamefont {Bending},
  \citenamefont {Como},\ and\ \citenamefont {van Heumen}}]{feng2023}%
  \BibitemOpen
  \bibfield  {author} {\bibinfo {author} {\bibfnamefont {X.}~\bibnamefont
  {Feng}}, \bibinfo {author} {\bibfnamefont {L.}~\bibnamefont {Farrar}},
  \bibinfo {author} {\bibfnamefont {C.~J.}\ \bibnamefont {Sayers}}, \bibinfo
  {author} {\bibfnamefont {S.~J.}\ \bibnamefont {Bending}}, \bibinfo {author}
  {\bibfnamefont {E.~D.}\ \bibnamefont {Como}},\ and\ \bibinfo {author}
  {\bibfnamefont {E.}~\bibnamefont {van Heumen}},\ }\bibfield  {title}
  {\bibinfo {title} {{Optical response of the bulk stabilized mosaic phase in
  Se doped TaS$_{2-x}$Se$_{x}$}},\ }\href {https://arxiv.org/abs/2311.15791}
  {\bibfield  {journal} {\bibinfo  {journal} {arXiv:2311.15791}\ } (\bibinfo
  {year} {2023})}\BibitemShut {NoStop}%
\bibitem [{\citenamefont {Gr\"uner}(1988)}]{Grunerreview1988}%
  \BibitemOpen
  \bibfield  {author} {\bibinfo {author} {\bibfnamefont {G.}~\bibnamefont
  {Gr\"uner}},\ }\bibfield  {title} {\bibinfo {title} {The dynamics of
  charge-density waves},\ }\href {https://doi.org/10.1103/RevModPhys.60.1129}
  {\bibfield  {journal} {\bibinfo  {journal} {Rev. Mod. Phys.}\ }\textbf
  {\bibinfo {volume} {60}},\ \bibinfo {pages} {1129} (\bibinfo {year}
  {1988})}\BibitemShut {NoStop}%
\bibitem [{\citenamefont {Pfuner}\ \emph {et~al.}(2010)\citenamefont {Pfuner},
  \citenamefont {Lerch}, \citenamefont {Chu}, \citenamefont {Kuo},
  \citenamefont {Fisher},\ and\ \citenamefont {Degiorgi}}]{MFPfuner2010}%
  \BibitemOpen
  \bibfield  {author} {\bibinfo {author} {\bibfnamefont {F.}~\bibnamefont
  {Pfuner}}, \bibinfo {author} {\bibfnamefont {P.}~\bibnamefont {Lerch}},
  \bibinfo {author} {\bibfnamefont {J.-H.}\ \bibnamefont {Chu}}, \bibinfo
  {author} {\bibfnamefont {H.-H.}\ \bibnamefont {Kuo}}, \bibinfo {author}
  {\bibfnamefont {I.~R.}\ \bibnamefont {Fisher}},\ and\ \bibinfo {author}
  {\bibfnamefont {L.}~\bibnamefont {Degiorgi}},\ }\bibfield  {title} {\bibinfo
  {title} {{Temperature dependence of the excitation spectrum in the
  charge-density-wave ${\text{ErTe}}_{3}$ and ${\text{HoTe}}_{3}$ systems}},\
  }\href {https://doi.org/10.1103/PhysRevB.81.195110} {\bibfield  {journal}
  {\bibinfo  {journal} {Phys. Rev. B}\ }\textbf {\bibinfo {volume} {81}},\
  \bibinfo {pages} {195110} (\bibinfo {year} {2010})}\BibitemShut {NoStop}%
\bibitem [{\citenamefont {Perucchi}\ \emph {et~al.}(2005)\citenamefont
  {Perucchi}, \citenamefont {Degiorgi},\ and\ \citenamefont
  {Berger}}]{Perucchi2005}%
  \BibitemOpen
  \bibfield  {author} {\bibinfo {author} {\bibfnamefont {A.}~\bibnamefont
  {Perucchi}}, \bibinfo {author} {\bibfnamefont {L.}~\bibnamefont {Degiorgi}},\
  and\ \bibinfo {author} {\bibfnamefont {H.}~\bibnamefont {Berger}},\
  }\bibfield  {title} {\bibinfo {title} {{Infrared signature of the
  charge-density-wave gap in ${\mathrm{ZrTe}}_{3}$}},\ }\href
  {https://doi.org/10.1140/epjb/e2006-00006-4} {\bibfield  {journal} {\bibinfo
  {journal} {Eur. Phys. J. B}\ }\textbf {\bibinfo {volume} {48}},\ \bibinfo
  {pages} {489} (\bibinfo {year} {2005})}\BibitemShut {NoStop}%
\bibitem [{\citenamefont {Uykur}\ \emph {et~al.}(2021)\citenamefont {Uykur},
  \citenamefont {Ortiz}, \citenamefont {Iakutkina}, \citenamefont {Wenzel},
  \citenamefont {Wilson}, \citenamefont {Dressel},\ and\ \citenamefont
  {Tsirlin}}]{EceCSV}%
  \BibitemOpen
  \bibfield  {author} {\bibinfo {author} {\bibfnamefont {E.}~\bibnamefont
  {Uykur}}, \bibinfo {author} {\bibfnamefont {B.~R.}\ \bibnamefont {Ortiz}},
  \bibinfo {author} {\bibfnamefont {O.}~\bibnamefont {Iakutkina}}, \bibinfo
  {author} {\bibfnamefont {M.}~\bibnamefont {Wenzel}}, \bibinfo {author}
  {\bibfnamefont {S.~D.}\ \bibnamefont {Wilson}}, \bibinfo {author}
  {\bibfnamefont {M.}~\bibnamefont {Dressel}},\ and\ \bibinfo {author}
  {\bibfnamefont {A.~A.}\ \bibnamefont {Tsirlin}},\ }\bibfield  {title}
  {\bibinfo {title} {{Low-energy optical properties of the nonmagnetic kagome
  metal ${\mathrm{CsV}}_{3}{\mathrm{Sb}}_{5}$}},\ }\href
  {https://doi.org/10.1103/PhysRevB.104.045130} {\bibfield  {journal} {\bibinfo
   {journal} {Phys. Rev. B}\ }\textbf {\bibinfo {volume} {104}},\ \bibinfo
  {pages} {045130} (\bibinfo {year} {2021})}\BibitemShut {NoStop}%
\bibitem [{\citenamefont {He}\ \emph {et~al.}(2024)\citenamefont {He},
  \citenamefont {Peis}, \citenamefont {Cuddy}, \citenamefont {Zhao},
  \citenamefont {Li}, \citenamefont {Zhang}, \citenamefont {Stumberger},
  \citenamefont {Moritz}, \citenamefont {Yang}, \citenamefont {Gao},
  \citenamefont {Devereaux},\ and\ \citenamefont {Hackl}}]{He2024}%
  \BibitemOpen
  \bibfield  {author} {\bibinfo {author} {\bibfnamefont {G.}~\bibnamefont
  {He}}, \bibinfo {author} {\bibfnamefont {L.}~\bibnamefont {Peis}}, \bibinfo
  {author} {\bibfnamefont {E.~F.}\ \bibnamefont {Cuddy}}, \bibinfo {author}
  {\bibfnamefont {Z.}~\bibnamefont {Zhao}}, \bibinfo {author} {\bibfnamefont
  {D.}~\bibnamefont {Li}}, \bibinfo {author} {\bibfnamefont {Y.}~\bibnamefont
  {Zhang}}, \bibinfo {author} {\bibfnamefont {R.}~\bibnamefont {Stumberger}},
  \bibinfo {author} {\bibfnamefont {B.}~\bibnamefont {Moritz}}, \bibinfo
  {author} {\bibfnamefont {H.}~\bibnamefont {Yang}}, \bibinfo {author}
  {\bibfnamefont {H.}~\bibnamefont {Gao}}, \bibinfo {author} {\bibfnamefont
  {T.~P.}\ \bibnamefont {Devereaux}},\ and\ \bibinfo {author} {\bibfnamefont
  {R.}~\bibnamefont {Hackl}},\ }\bibfield  {title} {\bibinfo {title}
  {{Anharmonic strong-coupling effects at the origin of the charge density wave
  in ${\mathrm{CsV}}_{3}{\mathrm{Sb}}_{5}$}},\ }\href
  {https://doi.org/10.1038/s41467-024-45865-0} {\bibfield  {journal} {\bibinfo
  {journal} {Nat. Commun.}\ }\textbf {\bibinfo {volume} {15}},\ \bibinfo
  {pages} {1895} (\bibinfo {year} {2024})}\BibitemShut {NoStop}%
\bibitem [{\citenamefont {Peierls}(1955)}]{Peierls1955}%
  \BibitemOpen
  \bibfield  {author} {\bibinfo {author} {\bibfnamefont {R.~F.}\ \bibnamefont
  {Peierls}},\ }\href@noop {} {\emph {\bibinfo {title} {Quantum Theory of
  Solids}}}\ (\bibinfo  {publisher} {Clarendon Press},\ \bibinfo {address}
  {Oxford},\ \bibinfo {year} {1955})\BibitemShut {NoStop}%
\bibitem [{\citenamefont {Kennedy}\ and\ \citenamefont
  {Lieb}(1987)}]{Kennedy1987}%
  \BibitemOpen
  \bibfield  {author} {\bibinfo {author} {\bibfnamefont {T.}~\bibnamefont
  {Kennedy}}\ and\ \bibinfo {author} {\bibfnamefont {E.~H.}\ \bibnamefont
  {Lieb}},\ }\bibfield  {title} {\bibinfo {title} {Proof of the peierls
  instability in one dimension},\ }\href
  {https://doi.org/10.1103/PhysRevLett.59.1309} {\bibfield  {journal} {\bibinfo
   {journal} {Phys. Rev. Lett.}\ }\textbf {\bibinfo {volume} {59}},\ \bibinfo
  {pages} {1309} (\bibinfo {year} {1987})}\BibitemShut {NoStop}%
\bibitem [{\citenamefont {Tanda}\ \emph {et~al.}(1984)\citenamefont {Tanda},
  \citenamefont {Sambongi}, \citenamefont {Tani},\ and\ \citenamefont
  {Tanaka}}]{X-ray1984}%
  \BibitemOpen
  \bibfield  {author} {\bibinfo {author} {\bibfnamefont {S.}~\bibnamefont
  {Tanda}}, \bibinfo {author} {\bibfnamefont {T.}~\bibnamefont {Sambongi}},
  \bibinfo {author} {\bibfnamefont {T.}~\bibnamefont {Tani}},\ and\ \bibinfo
  {author} {\bibfnamefont {S.}~\bibnamefont {Tanaka}},\ }\bibfield  {title}
  {\bibinfo {title} {{X-Ray Study of Charge Density Wave Structure in
  $1T$-${\mathrm{TaS}}_{2}$}},\ }\href {https://doi.org/10.1143/JPSJ.53.476}
  {\bibfield  {journal} {\bibinfo  {journal} {J. Phys. Soc. Jpn.}\ }\textbf
  {\bibinfo {volume} {53}},\ \bibinfo {pages} {476} (\bibinfo {year}
  {1984})}\BibitemShut {NoStop}%
\bibitem [{\citenamefont {Burri}\ \emph {et~al.}(2025)\citenamefont {Burri},
  \citenamefont {Bell}, \citenamefont {Dizdarević}, \citenamefont {Hu},
  \citenamefont {Ravnik}, \citenamefont {Vonka}, \citenamefont {Ekinci},
  \citenamefont {Huang}, \citenamefont {Gerber},\ and\ \citenamefont
  {Hua}}]{Burri2025}%
  \BibitemOpen
  \bibfield  {author} {\bibinfo {author} {\bibfnamefont {C.}~\bibnamefont
  {Burri}}, \bibinfo {author} {\bibfnamefont {H.~G.}\ \bibnamefont {Bell}},
  \bibinfo {author} {\bibfnamefont {F.}~\bibnamefont {Dizdarević}}, \bibinfo
  {author} {\bibfnamefont {W.}~\bibnamefont {Hu}}, \bibinfo {author}
  {\bibfnamefont {J.}~\bibnamefont {Ravnik}}, \bibinfo {author} {\bibfnamefont
  {J.}~\bibnamefont {Vonka}}, \bibinfo {author} {\bibfnamefont
  {Y.}~\bibnamefont {Ekinci}}, \bibinfo {author} {\bibfnamefont {S.-W.}\
  \bibnamefont {Huang}}, \bibinfo {author} {\bibfnamefont {S.}~\bibnamefont
  {Gerber}},\ and\ \bibinfo {author} {\bibfnamefont {N.}~\bibnamefont {Hua}},\
  }\bibfield  {title} {\bibinfo {title} {{Three-dimensional electronic domain
  correlations in $1T$-${\mathrm{TaS}}_{2}$}},\ }\href
  {https://arxiv.org/abs/2508.17839} {\bibfield  {journal} {\bibinfo  {journal}
  {arXiv:2508.17839}\ } (\bibinfo {year} {2025})}\BibitemShut {NoStop}%
\bibitem [{\citenamefont {Petkov}\ \emph {et~al.}(2022)\citenamefont {Petkov},
  \citenamefont {Peralta}, \citenamefont {Aoun},\ and\ \citenamefont
  {Ren}}]{Petkov2022}%
  \BibitemOpen
  \bibfield  {author} {\bibinfo {author} {\bibfnamefont {V.}~\bibnamefont
  {Petkov}}, \bibinfo {author} {\bibfnamefont {J.~E.}\ \bibnamefont {Peralta}},
  \bibinfo {author} {\bibfnamefont {B.}~\bibnamefont {Aoun}},\ and\ \bibinfo
  {author} {\bibfnamefont {Y.}~\bibnamefont {Ren}},\ }\bibfield  {title}
  {\bibinfo {title} {{Atomic structure and Mott nature of the insulating charge
  density wave phase of 1$T$-${\mathrm{TaS}}_{2}$}},\ }\href
  {https://doi.org/10.1088/1361-648X/ac77cf} {\bibfield  {journal} {\bibinfo
  {journal} {J. Phys. Condens. Matter}\ }\textbf {\bibinfo {volume} {34}},\
  \bibinfo {pages} {345401} (\bibinfo {year} {2022})}\BibitemShut {NoStop}%
\bibitem [{\citenamefont {Pal}\ \emph {et~al.}(2023)\citenamefont {Pal},
  \citenamefont {Bahera}, \citenamefont {Sahu}, \citenamefont {Srivastava},
  \citenamefont {Srivastava}, \citenamefont {Lalla}, \citenamefont {Sankar},
  \citenamefont {Banerjee},\ and\ \citenamefont {Roy}}]{PAL2023}%
  \BibitemOpen
  \bibfield  {author} {\bibinfo {author} {\bibfnamefont {S.}~\bibnamefont
  {Pal}}, \bibinfo {author} {\bibfnamefont {P.}~\bibnamefont {Bahera}},
  \bibinfo {author} {\bibfnamefont {S.}~\bibnamefont {Sahu}}, \bibinfo {author}
  {\bibfnamefont {H.}~\bibnamefont {Srivastava}}, \bibinfo {author}
  {\bibfnamefont {A.}~\bibnamefont {Srivastava}}, \bibinfo {author}
  {\bibfnamefont {N.}~\bibnamefont {Lalla}}, \bibinfo {author} {\bibfnamefont
  {R.}~\bibnamefont {Sankar}}, \bibinfo {author} {\bibfnamefont
  {A.}~\bibnamefont {Banerjee}},\ and\ \bibinfo {author} {\bibfnamefont
  {S.}~\bibnamefont {Roy}},\ }\bibfield  {title} {\bibinfo {title} {{Charge
  density wave and superconductivity in $6R$-${\mathrm{TaS}}_{2}$}},\ }\href
  {https://doi.org/https://doi.org/10.1016/j.physb.2023.415266} {\bibfield
  {journal} {\bibinfo  {journal} {Physica B: Condensed Matter}\ }\textbf
  {\bibinfo {volume} {669}},\ \bibinfo {pages} {415266} (\bibinfo {year}
  {2023})}\BibitemShut {NoStop}%
\bibitem [{\citenamefont {Yang}\ \emph {et~al.}(2024)\citenamefont {Yang},
  \citenamefont {Lee}, \citenamefont {Bang}, \citenamefont {Kim}, \citenamefont
  {Wulferding}, \citenamefont {Lee},\ and\ \citenamefont {Cho}}]{Origin2024}%
  \BibitemOpen
  \bibfield  {author} {\bibinfo {author} {\bibfnamefont {H.}~\bibnamefont
  {Yang}}, \bibinfo {author} {\bibfnamefont {B.}~\bibnamefont {Lee}}, \bibinfo
  {author} {\bibfnamefont {J.}~\bibnamefont {Bang}}, \bibinfo {author}
  {\bibfnamefont {S.}~\bibnamefont {Kim}}, \bibinfo {author} {\bibfnamefont
  {D.}~\bibnamefont {Wulferding}}, \bibinfo {author} {\bibfnamefont {S.-H.}\
  \bibnamefont {Lee}},\ and\ \bibinfo {author} {\bibfnamefont {D.}~\bibnamefont
  {Cho}},\ }\bibfield  {title} {\bibinfo {title} {{Origin of Distinct
  Insulating Domains in the Layered Charge Density Wave Material
  $1T$-${\mathrm{TaS}}_{2}$}},\ }\href {https://doi.org/10.1002/advs.202401348}
  {\bibfield  {journal} {\bibinfo  {journal} {Adv. Sci.}\ }\textbf {\bibinfo
  {volume} {11}},\ \bibinfo {pages} {2401348} (\bibinfo {year}
  {2024})}\BibitemShut {NoStop}%
\bibitem [{\citenamefont {Qazilbash}\ \emph {et~al.}(2009)\citenamefont
  {Qazilbash}, \citenamefont {Hamlin}, \citenamefont {Baumbach}, \citenamefont
  {Zhang}, \citenamefont {Singh}, \citenamefont {Maple},\ and\ \citenamefont
  {Basov}}]{Qazilbash2009}%
  \BibitemOpen
  \bibfield  {author} {\bibinfo {author} {\bibfnamefont {M.~M.}\ \bibnamefont
  {Qazilbash}}, \bibinfo {author} {\bibfnamefont {J.~J.}\ \bibnamefont
  {Hamlin}}, \bibinfo {author} {\bibfnamefont {R.~E.}\ \bibnamefont
  {Baumbach}}, \bibinfo {author} {\bibfnamefont {L.}~\bibnamefont {Zhang}},
  \bibinfo {author} {\bibfnamefont {D.~J.}\ \bibnamefont {Singh}}, \bibinfo
  {author} {\bibfnamefont {M.~B.}\ \bibnamefont {Maple}},\ and\ \bibinfo
  {author} {\bibfnamefont {D.~N.}\ \bibnamefont {Basov}},\ }\bibfield  {title}
  {\bibinfo {title} {Electronic correlations in the iron pnictides},\ }\href
  {https://doi.org/10.1038/nphys1343} {\bibfield  {journal} {\bibinfo
  {journal} {Nature Phys.}\ }\textbf {\bibinfo {volume} {5}},\ \bibinfo {pages}
  {647} (\bibinfo {year} {2009})}\BibitemShut {NoStop}%
\bibitem [{\citenamefont {Si}(2009)}]{Si2009}%
  \BibitemOpen
  \bibfield  {author} {\bibinfo {author} {\bibfnamefont {Q.}~\bibnamefont
  {Si}},\ }\bibfield  {title} {\bibinfo {title} {Electrons on the verge},\
  }\href {https://doi.org/10.1038/nphys1394} {\bibfield  {journal} {\bibinfo
  {journal} {Nature Phys.}\ }\textbf {\bibinfo {volume} {5}},\ \bibinfo {pages}
  {629} (\bibinfo {year} {2009})}\BibitemShut {NoStop}%
\bibitem [{\citenamefont {Ferber}\ \emph {et~al.}(2010)\citenamefont {Ferber},
  \citenamefont {Zhang}, \citenamefont {Jeschke},\ and\ \citenamefont
  {Valent\'{\i}}}]{Ferber2010}%
  \BibitemOpen
  \bibfield  {author} {\bibinfo {author} {\bibfnamefont {J.}~\bibnamefont
  {Ferber}}, \bibinfo {author} {\bibfnamefont {Y.-Z.}\ \bibnamefont {Zhang}},
  \bibinfo {author} {\bibfnamefont {H.~O.}\ \bibnamefont {Jeschke}},\ and\
  \bibinfo {author} {\bibfnamefont {R.}~\bibnamefont {Valent\'{\i}}},\
  }\bibfield  {title} {\bibinfo {title} {Analysis of spin-density wave
  conductivity spectra of iron pnictides in the framework of density functional
  theory},\ }\href {https://doi.org/10.1103/PhysRevB.82.165102} {\bibfield
  {journal} {\bibinfo  {journal} {Phys. Rev. B}\ }\textbf {\bibinfo {volume}
  {82}},\ \bibinfo {pages} {165102} (\bibinfo {year} {2010})}\BibitemShut
  {NoStop}%
\bibitem [{\citenamefont {Degiorgi}(2011)}]{Degiorgi2011}%
  \BibitemOpen
  \bibfield  {author} {\bibinfo {author} {\bibfnamefont {L.}~\bibnamefont
  {Degiorgi}},\ }\bibfield  {title} {\bibinfo {title} {Electronic correlations
  in iron-pnictide superconductors and beyond: lessons learned from optics},\
  }\href {https://doi.org/10.1088/1367-2630/13/2/023011} {\bibfield  {journal}
  {\bibinfo  {journal} {New J. Phys.}\ }\textbf {\bibinfo {volume} {13}},\
  \bibinfo {pages} {023011} (\bibinfo {year} {2011})}\BibitemShut {NoStop}%
\bibitem [{\citenamefont {Hovden}\ \emph {et~al.}(2016)\citenamefont {Hovden},
  \citenamefont {Tsen}, \citenamefont {Liu}, \citenamefont {Savitzky},
  \citenamefont {Baggari}, \citenamefont {Liu}, \citenamefont {Lu},
  \citenamefont {Sun}, \citenamefont {Kim}, \citenamefont {Pasupathy},\ and\
  \citenamefont {Kourkoutis}}]{Hovden2016}%
  \BibitemOpen
  \bibfield  {author} {\bibinfo {author} {\bibfnamefont {R.}~\bibnamefont
  {Hovden}}, \bibinfo {author} {\bibfnamefont {A.~W.}\ \bibnamefont {Tsen}},
  \bibinfo {author} {\bibfnamefont {P.}~\bibnamefont {Liu}}, \bibinfo {author}
  {\bibfnamefont {B.~H.}\ \bibnamefont {Savitzky}}, \bibinfo {author}
  {\bibfnamefont {I.~E.}\ \bibnamefont {Baggari}}, \bibinfo {author}
  {\bibfnamefont {Y.}~\bibnamefont {Liu}}, \bibinfo {author} {\bibfnamefont
  {W.}~\bibnamefont {Lu}}, \bibinfo {author} {\bibfnamefont {Y.}~\bibnamefont
  {Sun}}, \bibinfo {author} {\bibfnamefont {P.}~\bibnamefont {Kim}}, \bibinfo
  {author} {\bibfnamefont {A.~N.}\ \bibnamefont {Pasupathy}},\ and\ \bibinfo
  {author} {\bibfnamefont {L.~F.}\ \bibnamefont {Kourkoutis}},\ }\bibfield
  {title} {\bibinfo {title} {{Atomic lattice disorder in charge-density-wave
  phases of exfoliated dichalcogenides ($1T$-${\mathrm{TaS}}_{2}$)}},\ }\href
  {https://doi.org/10.1073/pnas.1606044113} {\bibfield  {journal} {\bibinfo
  {journal} {Proc. Natl Acad. Sci. USA}\ }\textbf {\bibinfo {volume} {113}},\
  \bibinfo {pages} {11420} (\bibinfo {year} {2016})}\BibitemShut {NoStop}%
\bibitem [{\citenamefont {Stahl}\ \emph {et~al.}(2020)\citenamefont {Stahl},
  \citenamefont {Kusch}, \citenamefont {Heinsch}, \citenamefont {Garbarino},
  \citenamefont {Kretzschmar}, \citenamefont {Hanff}, \citenamefont
  {Rossnagel}, \citenamefont {Geck},\ and\ \citenamefont
  {Ritschel}}]{Stahl2020}%
  \BibitemOpen
  \bibfield  {author} {\bibinfo {author} {\bibfnamefont {Q.}~\bibnamefont
  {Stahl}}, \bibinfo {author} {\bibfnamefont {M.}~\bibnamefont {Kusch}},
  \bibinfo {author} {\bibfnamefont {F.}~\bibnamefont {Heinsch}}, \bibinfo
  {author} {\bibfnamefont {G.}~\bibnamefont {Garbarino}}, \bibinfo {author}
  {\bibfnamefont {N.}~\bibnamefont {Kretzschmar}}, \bibinfo {author}
  {\bibfnamefont {K.}~\bibnamefont {Hanff}}, \bibinfo {author} {\bibfnamefont
  {K.}~\bibnamefont {Rossnagel}}, \bibinfo {author} {\bibfnamefont
  {J.}~\bibnamefont {Geck}},\ and\ \bibinfo {author} {\bibfnamefont
  {T.}~\bibnamefont {Ritschel}},\ }\bibfield  {title} {\bibinfo {title}
  {{Collapse of layer dimerization in the photo-induced hidden state of
  $1T$-${\mathrm{TaS}}_{2}$}},\ }\href
  {https://doi.org/10.1038/s41467-020-15079-1} {\bibfield  {journal} {\bibinfo
  {journal} {Nat. Commun.}\ }\textbf {\bibinfo {volume} {11}},\ \bibinfo
  {pages} {1247} (\bibinfo {year} {2020})}\BibitemShut {NoStop}%
\bibitem [{\citenamefont {Liu}\ \emph {et~al.}(2026)\citenamefont {Liu},
  \citenamefont {Liu}, \citenamefont {Yang}, \citenamefont {Lee}, \citenamefont
  {Pan}, \citenamefont {Chen}, \citenamefont {Huang}, \citenamefont {Jiang},
  \citenamefont {Hu}, \citenamefont {Zhang}, \citenamefont {Xie}, \citenamefont
  {Wang}, \citenamefont {Guan}, \citenamefont {Jiang}, \citenamefont {Yang},
  \citenamefont {Li}, \citenamefont {Yun}, \citenamefont {Wang}, \citenamefont
  {Meng}, \citenamefont {Yao}, \citenamefont {Qian},\ and\ \citenamefont
  {Shi}}]{Liu2025}%
  \BibitemOpen
  \bibfield  {author} {\bibinfo {author} {\bibfnamefont {J.}~\bibnamefont
  {Liu}}, \bibinfo {author} {\bibfnamefont {P.}~\bibnamefont {Liu}}, \bibinfo
  {author} {\bibfnamefont {L.}~\bibnamefont {Yang}}, \bibinfo {author}
  {\bibfnamefont {S.-H.}\ \bibnamefont {Lee}}, \bibinfo {author} {\bibfnamefont
  {M.}~\bibnamefont {Pan}}, \bibinfo {author} {\bibfnamefont {F.}~\bibnamefont
  {Chen}}, \bibinfo {author} {\bibfnamefont {J.}~\bibnamefont {Huang}},
  \bibinfo {author} {\bibfnamefont {B.}~\bibnamefont {Jiang}}, \bibinfo
  {author} {\bibfnamefont {M.}~\bibnamefont {Hu}}, \bibinfo {author}
  {\bibfnamefont {Y.}~\bibnamefont {Zhang}}, \bibinfo {author} {\bibfnamefont
  {Z.}~\bibnamefont {Xie}}, \bibinfo {author} {\bibfnamefont {G.}~\bibnamefont
  {Wang}}, \bibinfo {author} {\bibfnamefont {M.}~\bibnamefont {Guan}}, \bibinfo
  {author} {\bibfnamefont {W.}~\bibnamefont {Jiang}}, \bibinfo {author}
  {\bibfnamefont {H.}~\bibnamefont {Yang}}, \bibinfo {author} {\bibfnamefont
  {J.}~\bibnamefont {Li}}, \bibinfo {author} {\bibfnamefont {C.}~\bibnamefont
  {Yun}}, \bibinfo {author} {\bibfnamefont {Z.}~\bibnamefont {Wang}}, \bibinfo
  {author} {\bibfnamefont {S.}~\bibnamefont {Meng}}, \bibinfo {author}
  {\bibfnamefont {Y.}~\bibnamefont {Yao}}, \bibinfo {author} {\bibfnamefont
  {T.}~\bibnamefont {Qian}},\ and\ \bibinfo {author} {\bibfnamefont
  {X.}~\bibnamefont {Shi}},\ }\bibfield  {title} {\bibinfo {title}
  {{Nonvolatile optical control of interlayer stacking order in
  $1T$-${\mathrm{TaS}}_{2}$}},\ }\href
  {https://doi.org/10.1038/s41535-025-00836-6} {\bibfield  {journal} {\bibinfo
  {journal} {npj Quantum Mater.}\ }\textbf {\bibinfo {volume} {11}},\ \bibinfo
  {pages} {6} (\bibinfo {year} {2026})}\BibitemShut {NoStop}%
\bibitem [{\citenamefont {Bae}\ \emph {et~al.}(2025)\citenamefont {Bae},
  \citenamefont {Valent\'{\i}}, \citenamefont {Mazin},\ and\ \citenamefont
  {Yan}}]{Bae2025}%
  \BibitemOpen
  \bibfield  {author} {\bibinfo {author} {\bibfnamefont {H.}~\bibnamefont
  {Bae}}, \bibinfo {author} {\bibfnamefont {R.}~\bibnamefont {Valent\'{\i}}},
  \bibinfo {author} {\bibfnamefont {I.~I.}\ \bibnamefont {Mazin}},\ and\
  \bibinfo {author} {\bibfnamefont {B.}~\bibnamefont {Yan}},\ }\bibfield
  {title} {\bibinfo {title} {{Designing flat bands, localized and itinerant
  states in ${\mathrm{TaS}}_{2}$ trilayer heterostructures}},\ }\href
  {https://doi.org/10.1038/s41535-025-00812-0} {\bibfield  {journal} {\bibinfo
  {journal} {npj Quantum Mater.}\ }\textbf {\bibinfo {volume} {10}},\ \bibinfo
  {pages} {92} (\bibinfo {year} {2025})}\BibitemShut {NoStop}%
\bibitem [{\citenamefont {Tanatar}\ \emph {et~al.}(2010)\citenamefont
  {Tanatar}, \citenamefont {Ni}, \citenamefont {Thaler}, \citenamefont
  {Bud'ko}, \citenamefont {Canfield},\ and\ \citenamefont
  {Prozorov}}]{Tantar2010}%
  \BibitemOpen
  \bibfield  {author} {\bibinfo {author} {\bibfnamefont {M.~A.}\ \bibnamefont
  {Tanatar}}, \bibinfo {author} {\bibfnamefont {N.}~\bibnamefont {Ni}},
  \bibinfo {author} {\bibfnamefont {A.}~\bibnamefont {Thaler}}, \bibinfo
  {author} {\bibfnamefont {S.~L.}\ \bibnamefont {Bud'ko}}, \bibinfo {author}
  {\bibfnamefont {P.~C.}\ \bibnamefont {Canfield}},\ and\ \bibinfo {author}
  {\bibfnamefont {R.}~\bibnamefont {Prozorov}},\ }\bibfield  {title} {\bibinfo
  {title} {{Pseudogap and its critical point in the heavily doped
  $\text{Ba}{({\text{Fe}}_{1\ensuremath{-}x}{\text{Co}}_{x})}_{2}{\text{As}}_{2}$
  from $c$-axis resistivity measurements}},\ }\href
  {https://doi.org/10.1103/PhysRevB.82.134528} {\bibfield  {journal} {\bibinfo
  {journal} {Phys. Rev. B}\ }\textbf {\bibinfo {volume} {82}},\ \bibinfo
  {pages} {134528} (\bibinfo {year} {2010})}\BibitemShut {NoStop}%
\bibitem [{\citenamefont {Dressel}\ and\ \citenamefont
  {Gr\"uner}(2002)}]{Dressel2002}%
  \BibitemOpen
  \bibfield  {author} {\bibinfo {author} {\bibfnamefont {M.}~\bibnamefont
  {Dressel}}\ and\ \bibinfo {author} {\bibfnamefont {G.}~\bibnamefont
  {Gr\"uner}},\ }\href@noop {} {\emph {\bibinfo {title} {Electrodynamics of
  Solids: Optical Properties of Electrons in Matter}}}\ (\bibinfo  {publisher}
  {Cambridge University Press},\ \bibinfo {address} {Cambridge},\ \bibinfo
  {year} {2002})\BibitemShut {NoStop}%
\bibitem [{\citenamefont {Giannozzi}\ \emph {et~al.}(2009)\citenamefont
  {Giannozzi}, \citenamefont {Baroni}, \citenamefont {Bonini}, \citenamefont
  {Calandra}, \citenamefont {Car}, \citenamefont {Cavazzoni}, \citenamefont
  {Ceresoli}, \citenamefont {Chiarotti}, \citenamefont {Cococcioni},
  \citenamefont {Dabo}, \citenamefont {Dal~Corso},\ and\ \citenamefont
  {de~Gironcoli}}]{Giannozzi2009}%
  \BibitemOpen
  \bibfield  {author} {\bibinfo {author} {\bibfnamefont {P.}~\bibnamefont
  {Giannozzi}}, \bibinfo {author} {\bibfnamefont {S.}~\bibnamefont {Baroni}},
  \bibinfo {author} {\bibfnamefont {N.}~\bibnamefont {Bonini}}, \bibinfo
  {author} {\bibfnamefont {M.}~\bibnamefont {Calandra}}, \bibinfo {author}
  {\bibfnamefont {R.}~\bibnamefont {Car}}, \bibinfo {author} {\bibfnamefont
  {C.}~\bibnamefont {Cavazzoni}}, \bibinfo {author} {\bibfnamefont
  {D.}~\bibnamefont {Ceresoli}}, \bibinfo {author} {\bibfnamefont {G.~L.}\
  \bibnamefont {Chiarotti}}, \bibinfo {author} {\bibfnamefont {M.}~\bibnamefont
  {Cococcioni}}, \bibinfo {author} {\bibfnamefont {I.}~\bibnamefont {Dabo}},
  \bibinfo {author} {\bibfnamefont {A.}~\bibnamefont {Dal~Corso}},\ and\
  \bibinfo {author} {\bibfnamefont {S.~e.~a.}\ \bibnamefont {de~Gironcoli}},\
  }\bibfield  {title} {\bibinfo {title} {{QUANTUM ESPRESSO: a modular and
  open-source software project for quantum simulations of materials}},\ }\href
  {https://doi.org/10.1088/0953-8984/21/39/395502} {\bibfield  {journal}
  {\bibinfo  {journal} {J. Phys. Condens. Matter}\ }\textbf {\bibinfo {volume}
  {21}},\ \bibinfo {pages} {395502} (\bibinfo {year} {2009})}\BibitemShut
  {NoStop}%
\bibitem [{\citenamefont {Giannozzi}\ \emph {et~al.}(2017)\citenamefont
  {Giannozzi}, \citenamefont {Andreussi}, \citenamefont {Brumme}, \citenamefont
  {Bunau}, \citenamefont {Buongiorno~Nardelli}, \citenamefont {Calandra},
  \citenamefont {Car}, \citenamefont {Cavazzoni}, \citenamefont {Ceresoli},
  \citenamefont {Cococcioni}, \citenamefont {Colonna},\ and\ \citenamefont
  {Carnimeo}}]{Giannozzi2017}%
  \BibitemOpen
  \bibfield  {author} {\bibinfo {author} {\bibfnamefont {P.}~\bibnamefont
  {Giannozzi}}, \bibinfo {author} {\bibfnamefont {O.}~\bibnamefont
  {Andreussi}}, \bibinfo {author} {\bibfnamefont {T.}~\bibnamefont {Brumme}},
  \bibinfo {author} {\bibfnamefont {O.}~\bibnamefont {Bunau}}, \bibinfo
  {author} {\bibfnamefont {M.}~\bibnamefont {Buongiorno~Nardelli}}, \bibinfo
  {author} {\bibfnamefont {M.}~\bibnamefont {Calandra}}, \bibinfo {author}
  {\bibfnamefont {R.}~\bibnamefont {Car}}, \bibinfo {author} {\bibfnamefont
  {C.}~\bibnamefont {Cavazzoni}}, \bibinfo {author} {\bibfnamefont
  {D.}~\bibnamefont {Ceresoli}}, \bibinfo {author} {\bibfnamefont
  {M.}~\bibnamefont {Cococcioni}}, \bibinfo {author} {\bibfnamefont
  {N.}~\bibnamefont {Colonna}},\ and\ \bibinfo {author} {\bibfnamefont
  {I.~e.~a.}\ \bibnamefont {Carnimeo}},\ }\bibfield  {title} {\bibinfo {title}
  {{Advanced capabilities for materials modelling with Quantum ESPRESSO}},\
  }\href {https://doi.org/10.1088/1361-648X/aa8f79} {\bibfield  {journal}
  {\bibinfo  {journal} {J. Phys. Condens. Matter}\ }\textbf {\bibinfo {volume}
  {29}},\ \bibinfo {pages} {465901} (\bibinfo {year} {2017})}\BibitemShut
  {NoStop}%
\bibitem [{\citenamefont {Blaha}\ \emph {et~al.}()\citenamefont {Blaha},
  \citenamefont {Schwarz}, \citenamefont {Madsen}, \citenamefont {Kvasnicka},
  \citenamefont {Luitz}, \citenamefont {Laskowski}, \citenamefont {Tran},\ and\
  \citenamefont {Marks}}]{wien2k}%
  \BibitemOpen
  \bibfield  {author} {\bibinfo {author} {\bibfnamefont {P.}~\bibnamefont
  {Blaha}}, \bibinfo {author} {\bibfnamefont {K.}~\bibnamefont {Schwarz}},
  \bibinfo {author} {\bibfnamefont {G.}~\bibnamefont {Madsen}}, \bibinfo
  {author} {\bibfnamefont {D.}~\bibnamefont {Kvasnicka}}, \bibinfo {author}
  {\bibfnamefont {J.}~\bibnamefont {Luitz}}, \bibinfo {author} {\bibfnamefont
  {R.}~\bibnamefont {Laskowski}}, \bibinfo {author} {\bibfnamefont
  {F.}~\bibnamefont {Tran}},\ and\ \bibinfo {author} {\bibfnamefont
  {L.}~\bibnamefont {Marks}},\ }\href@noop {} {}\bibinfo {note} {WIEN2k, An
  Augmented Plane Wave + Local Orbitals Program for Calculating Crystal
  Properties (Karlheinz Schwarz, Techn. Universit\"at Wien, Austria), 2018.
  ISBN 3-9501031-1-2}\BibitemShut {NoStop}%
\bibitem [{\citenamefont {Blaha}\ \emph {et~al.}(2020)\citenamefont {Blaha},
  \citenamefont {Schwarz}, \citenamefont {Tran}, \citenamefont {Laskowski},
  \citenamefont {Madsen},\ and\ \citenamefont {Marks}}]{Blaha2020}%
  \BibitemOpen
  \bibfield  {author} {\bibinfo {author} {\bibfnamefont {P.}~\bibnamefont
  {Blaha}}, \bibinfo {author} {\bibfnamefont {K.}~\bibnamefont {Schwarz}},
  \bibinfo {author} {\bibfnamefont {F.}~\bibnamefont {Tran}}, \bibinfo {author}
  {\bibfnamefont {R.}~\bibnamefont {Laskowski}}, \bibinfo {author}
  {\bibfnamefont {G.~K.~H.}\ \bibnamefont {Madsen}},\ and\ \bibinfo {author}
  {\bibfnamefont {L.~D.}\ \bibnamefont {Marks}},\ }\bibfield  {title} {\bibinfo
  {title} {{WIEN2k: An APW+lo program for calculating the properties of
  solids}},\ }\href {https://doi.org/10.1063/1.5143061} {\bibfield  {journal}
  {\bibinfo  {journal} {J. Chem. Phys.}\ }\textbf {\bibinfo {volume} {152}},\
  \bibinfo {pages} {074101} (\bibinfo {year} {2020})}\BibitemShut {NoStop}%
\bibitem [{\citenamefont {Ambrosch-Draxl}\ and\ \citenamefont
  {Sofo}(2006)}]{Draxl2006}%
  \BibitemOpen
  \bibfield  {author} {\bibinfo {author} {\bibfnamefont {C.}~\bibnamefont
  {Ambrosch-Draxl}}\ and\ \bibinfo {author} {\bibfnamefont {J.~O.}\
  \bibnamefont {Sofo}},\ }\bibfield  {title} {\bibinfo {title} {{Linear optical
  properties of solids within the full-potential linearized augmented planewave
  method}},\ }\href {https://doi.org/https://doi.org/10.1016/j.cpc.2006.03.005}
  {\bibfield  {journal} {\bibinfo  {journal} {Comput. Phys. Commun.}\ }\textbf
  {\bibinfo {volume} {175}},\ \bibinfo {pages} {1} (\bibinfo {year}
  {2006})}\BibitemShut {NoStop}%
\bibitem [{\citenamefont {Suh}\ \emph {et~al.}(2000)\citenamefont {Suh},
  \citenamefont {Park},\ and\ \citenamefont {Kim}}]{Suh2000}%
  \BibitemOpen
  \bibfield  {author} {\bibinfo {author} {\bibfnamefont {I.-H.}\ \bibnamefont
  {Suh}}, \bibinfo {author} {\bibfnamefont {Y.-S.}\ \bibnamefont {Park}},\ and\
  \bibinfo {author} {\bibfnamefont {J.-G.}\ \bibnamefont {Kim}},\ }\bibfield
  {title} {\bibinfo {title} {{{\it ORTHON}: transformation from triclinic axes
  and atomic coordinates to orthonormal ones}},\ }\href
  {https://doi.org/10.1107/S0021889800006579} {\bibfield  {journal} {\bibinfo
  {journal} {J. Appl. Crystallogr.}\ }\textbf {\bibinfo {volume} {33}},\
  \bibinfo {pages} {994} (\bibinfo {year} {2000})}\BibitemShut {NoStop}%
\bibitem [{\citenamefont {Persson}(2014)}]{osti_1192226}%
  \BibitemOpen
  \bibfield  {author} {\bibinfo {author} {\bibfnamefont {K.}~\bibnamefont
  {Persson}},\ }\href {https://doi.org/10.17188/1192226} {\bibinfo {title}
  {{Materials Data on TaS$_2$ (SG:164) by Materials Project}}} (\bibinfo {year}
  {2014})\BibitemShut {NoStop}%
\end{thebibliography}
\end{document}